\documentclass[twocolumn]{aastex631}

\usepackage{amsmath}
\begin{document}

\title{Constraints on the galaxy formation models during epoch of reionization with high redshift observations}
\correspondingauthor{Qing-Bo Ma}
\email{maqb@gznu.edu.cn}
\author[0000-0001-9493-4565]{Qing-Bo Ma}
\affil{School of Physics and Electronic Science, Guizhou Normal University, Guiyang 550001, PR China}
\affil{Guizhou Provincial Key Laboratory of Radio Astronomy and Data Processing, \\
Guizhou Normal University, Guiyang 550001, PR China}

\author{Xiao-Rong Chen}
\affil{School of Physics and Electronic Science, Guizhou Normal University, Guiyang 550001, PR China}

\author[0000-0002-1318-4828]{Ming Li}
\affil{Key Laboratory for Computational Astrophysics, National Astronomical Observatories, \\
Chinese Academy of Sciences, Beijing 100101, China}

\author{Qi Guo}
\affil{Key Laboratory for Computational Astrophysics, National Astronomical Observatories, \\ 
Chinese Academy of Sciences, Beijing 100101, China}
\affil{Institute for Frontiers in Astronomy and Astrophysics, Beijing Normal University, Beijing 102206, China}
\affil{School of Astronomy and Space Science, University of Chinese Academy of Sciences, Beijing 100049, China}

\author{Benedetta Ciardi}
\affil{Max-Planck-Institut für Astrophysik, Garching 85748, Germany}

\author[0000-0003-3401-4884]{Anshuman Acharya}
\affil{Max-Planck-Institut für Astrophysik, Garching 85748, Germany}

\author[0000-0002-9373-3865]{Xin Wang}
\affil{School of Astronomy and Space Science, University of Chinese Academy of Sciences, Beijing 100049, China}
\affil{National Astronomical Observatories, Chinese Academy of Sciences, Beijing 100101, China}
\affil{Institute for Frontiers in Astronomy and Astrophysics, Beijing Normal University, Beijing 102206, China}

\begin{abstract}
We use high resolution {\it N}-body dark matter simulations and {\sc L-Galaxies} semi-analytical galaxy formation models to explore the high-$z$ galaxy properties and estimate the budget of ionizing photons.
The parameters within {\sc L-Galaxies} are obtained using a Markov Chain Monte Carlo (MCMC) method with high-$z$ galaxy observations from JWST and other telescopes.
We consider two versions of {\sc L-Galaxies} with and without dust correction on galaxy UV luminosities. 
With the best-fit parameters, both {\sc L-Galaxies 2015} and {\sc L-Galaxies 2020} reproduce well observations of UV luminosity functions, stellar mass functions, star formation rate densities and ionizing photon emission efficiency.
With the assumption of escape fraction of $20\%$, all models produce more ionizing photons than the number of Hydrogen atoms in the Universe at $z>6$.
The inclusion of dust correction within MCMC results in higher star formation efficiency, which predicts $\sim 50\%$ more ionizing photons, with better consistency between the predicted stellar mass functions and observations.
\end{abstract}

\keywords{dark ages, reionization, first stars - galaxies: formation - methods: numerical}

\section{Introduction} 
\label{sec:intro}
The epoch of reionization (EoR) is the phase transition process of the Universe from cold and neutral to hot and ionized, driven by stellar and galactic radiation sources \citep{Furlanetto2006}. 
Observations of Gunn-Peterson (GP) troughs in QSO spectra suggest that the EoR should have been completed by $z\sim 6$ \citep{Fan2006},  
or even at lower $z$ \cite[e.g. ][]{Becker2015, Bosman2022}. 
Cosmic Microwave Background (CMB) experiments \cite[e.g. the Planck project, ][]{Planck2020A&A} measured a Thomson scattering optical depth $\tau = 0.0561 \pm 0.0071$,  which, for the parametrization  of the reionization history adopted in \citet{Planck2020A&A}, results in a mid-point of reionization of $z=7.82 \pm 0.71$.
Many state-of-the-art astronomical facilities have been recently providing superb data on celestial objects during the EoR. Among these, early results from the James Webb Space Telescope (JWST) suggested the presence of galaxies at redshifts as high as  $z=16$ \citep{Harikane2023ApJS}, while a galaxy at $z=14$ has been spectroscopically confirmed  \citep{Carniani2024Natur}. 
The Atacama Large Millimeter/submillimeter Array (ALMA) telescope has measured the properties of galaxies at $z>6$ through e.g. the [CII] emission line \citep{Bouwens2020}.
Radio telescopes such as the low-frequency array (LOFAR), the Murchison Widefield Array (MWA), and the Hydrogen Epoch of Reionization Array (HERA) have already released some upper limit observations of the power spectra of the 21-cm line from neutral hydrogen \citep{Mertens2020, Acharya2024MNRAS, Trott2020, Abdurashidova2022}, while the next generation telescope Square Kilometre Array (SKA) will provide a more detailed picture of the history and topology of the EoR  \citep{Koopmans2015}. 

As galaxies are believed to be the main ionizing sources during the EoR, investigating high-$z$ galaxies can help to understand the reionization process \citep{Dayal2018}. 
In the last few years, the JWST telescope has observed many low-mass and faint high-$z$ galaxies, due to its depth and wavelength coverage, which allows to study the properties of very low-mass galaxies  such as  stellar mass and SFR  during the EoR \citep{Navarro-Carrera2024, Wang2024ApJ}.
Meanwhile, the JWST has measured the rest-frame UV luminosity functions (UVLF) of galaxies up to $z=16$ \citep{Harikane2023ApJS, Adams2024ApJ, Donnan2024MNRAS, Finkelstein2024ApJ}.
The Hubble Space Telescope (HST) and the Spitzer telescope also measured the UVLF and stellar mass functions (SMF) of galaxies at $z>6$ \citep{Bouwens2021, Stefanon2021}.
Note that these results are mostly from observed galaxies with photometric confirmed redshifts, while the UVLF with spectroscopic redshifts are still with very low precision \citep{Harikane2024ApJ, Harikane2024arXiv}, since only a small fraction of high-$z$ galaxies is spectroscopically confirmed.

Ideally, galaxy formation and evolution should be consistently modelled with the reionization process through high-resolution radiation hydrodynamical simulations. 
Although several are present in the literature, e.g. FIRE \citep{MaX2018}, SPHINX \citep{Rosdahl2018}, CROC \citep{Esmerian2021},  THESAN \citep{Kannan2022}, CoDa \citep{Ocvirk2020MNRAS} and SPICE \citep{Bhagwat2024MNRAS}, they are extremely computationally expensive, so that they are necessarily limited in terms of e.g. box dimension and/or parameter space exploration.
More efficient semi-analytical/numerical approaches which follow the formation and evolution of galaxies within the reionization process have been developed, among which MERAXES \citep{Mutch2016}, ASTRAEUS \citep{Hutter2021}, and POLAR \citep{Ma2023MNRAS, Acharya2024arXiv}.
These approaches usually adopt the halo distribution and merger tree from {\it N}-body dark matter simulations, and model the formation and evolution of galaxies according to some prescription describing the various physical processes at play, e.g. star formation, metal enrichment, feedback effects etc.
Although these approaches are extremely efficient, they are still very expensive to fit the model parameters with  observations, employing e.g. a Markov Chain Monte Carlo (MCMC) method.
To make the MCMC process more efficient, \cite{Henriques2013} proposed to randomly select merger trees and halo samples to produce statistically less precise (e.g. within $\sim 5\%$ errors) SMF and luminosity functions before the MCMC runs. 
This technique is applied in \cite{Henriques_2015, Henriques_2020} to find galaxy formation parameters which reproduce low-$z$ observations.
The analytic models that are based on  halo mass functions are usually very efficient in computing the UVLF and SMF. Such methods have already been employed in MCMC calculations \cite[e.g. ][]{Park2019MNRAS, Zhang2022MNRAS}, but they do not naturally include e.g. the  stochasticity of the UV luminosity versus halo mass relation \citep{Gelli2024ApJ, Nikolic2024arXiv}, nor a more physically motivated description of the galaxy formation and evolution model. 

In this paper, we will apply the MCMC technique developed in \cite{Henriques2013} to explore how high-$z$ galaxy observations from the JWST telescope constrain galaxy formation models, by combining high resolution {\it N}-body dark matter simulations with the semi-analytical galaxy formation model (SAM) {\sc L-Galaxies} \citep{Henriques_2015, Henriques_2020}.
We will then explore the galaxy properties and estimate the budget of ionizing photons with the best-fit parameters from different sets of MCMC runs. 

The paper is organized as follows: we describe the {\it N}-body dark matter simulations, galaxy formation models and MCMC calculations in Section \ref{sec:simul}, the results of high-$z$ galaxy properties and budget of ionizing photons are presented in Section \ref{sec:resul}, and the conclusions are summarized in Section \ref{sec:conclu}.
The cosmological parameters adopted for the {\it N}-body dark matter and {\sc L-Galaxies} simulations are from the {\it Planck} project \citep{Planck2020A&A} fitted with the data sets of TT, TE, and EE+lowE+lensing+BAO, i.e. $\Omega_{\Lambda}= 0.6889$, $\Omega_{m} = 0.3111$, $\Omega_{b} = 0.049$, $h = 0.6766$, $\sigma_{8} = 0.8102$ and $n_{s} = 0.9665$.

\section{Methods} 
\label{sec:simul}

\subsection{Dark matter simulations}
For the distributions and the merger trees of dark matter halos we adopt the {\sc Jiutian-300} {\it N}-body dark matter simulation. 
It was run with the {\sc Gadget-4} code \citep{Springel2021}, with a box size of $300\,{\rm cMpc}/h$ and  $6144^{3}$ dark matter particles, corresponding to a particle mass resolution of $1.0 \times 10^{7}\,{\rm M_{\odot}}/h$.
The simulations start at $z=127$ and end at $z=0$, outputting a total number of 128 snapshots, although in this paper we only employ the first 39 with redshift $z \ge 6$.
The Friend-of-Friend (FoF) algorithm \citep{Springel2001} is applied to define dark matter halos, while the Subfind technique is used to identify sub-halos. 
The halos have at least 20 dark-matter particles, i.e. the minimum halo mass is $2.0 \times 10^8\,{\rm M_{\odot}}/h$. 
They are then used to construct the halo merger trees by following \cite{Springel2005}.
\begin{figure}
\centering
	\includegraphics[width=1.0\linewidth]{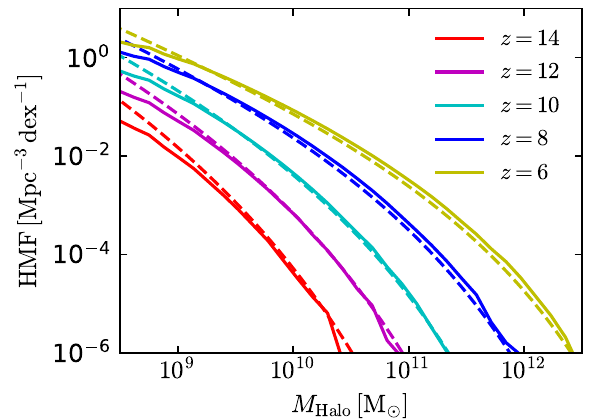}
    \caption{Halo mass functions (HMF) at $z=14$ (red), 12 (magenta), 10 (cyan), 8 (blue) and 6 (yellow). Solid lines refer to results from the {\sc Jiutian-300} simulation, while dashed lines are those from \citet{Tinker2008}, computed with the COLIBRI library. 
    }
    \label{fig:hmf_jiutian}
\end{figure}
As a reference, in Fig.~\ref{fig:hmf_jiutian} we show the halo mass function (HMF) at five redshifts, from 14 to 6.
As a comparison, we also plot the HMF from \cite{Tinker2008}, which are computed with the COLIBRI\footnote{https://github.com/GabrieleParimbelli/COLIBRI} library and are roughly consistent with those from our simulations, although some differences are observed at the low mass end, probably due to the resolution which is not enough to properly capture halos of those masses.

\subsection{{\sc L-Galaxies} models}
As our fiducial semi-analytical galaxy formation model (SAM), we adopt {\sc L-Galaxies 2020} \cite[named LG20 in the following;][]{Henriques_2020}, which is an updated version of {\sc L-Galaxies 2015} \cite[named LG15 in the following;][]{Henriques_2015}.
Both LG15 and LG20 include many physical processes related to galaxy formation and evolution, e.g. gas cooling, star formation, galaxy merger, supernovae feedback, black hole growth and AGN feedback, chemical enrichment, stellar population synthesis (SPS) models, tidal effects, and reincorporation of ejected gas.
Compared to LG15, LG20 adds a molecular hydrogen (H$_{2}$) based star formation prescription, spatially-resolved rings of the gas and stellar disc in galaxies, and gas radial flow processes. 
Differently, LG15 adopts the standard star formation prescription from \cite{Croton2006MNRAS} and \cite{Guo2011MNRAS}, in which the star formation rate (SFR) is linearly related to the cold gas mass within galaxies when the gas surface density is above a given mass threshold.
For more comparisons of different L-Galaxies versions, one can refer to the recent paper \cite{Vani2024}. 

Galaxy properties such as stellar mass, SFR and metal enrichment are directly computed in both LG15 and LG20, and also the UV luminosity and ionizing photon budget are calculated on the fly by adopting spectral energy distributions (SED) of the binary star mode from Binary Population and Spectral Synthesis \cite[BPASS, ][]{Eldridge2017, Stanway2018MNRAS}. Note that choosing a different SPS model can result in a very different ionizing photon budget, while this does not strongly affect the UV luminosities \citep{Liu2024ApJ}. 
We also note that, differently from LG15, LG20 includes a time delay after star formation for a supernovae explosion to happen, which results in a delayed energy and metal ejection in comparison to LG15.

\subsection{MCMC calculation}
\label{sec:mcmc_cal}
There are many free parameters in the modeling of physical processes related to galaxy formation and evolution.
\cite{Henriques2013} employed a Markov Chain Monte Carlo (MCMC) technique to constrain the values of these parameters in order to match the results from {\sc L-Galaxies} simulations (named LG in the following) to observations. 
Following \cite{Henriques2013}, we firstly apply the fiducial LG20 results to select the random tree samples. 
Note that, the fiducial LG20 results are only used to check if galaxy properties such as SMF and UVLF obtained from random tree samples are still consistent with the LG results. 
To make the MCMC more efficient, we do not select the biggest merger trees.
Besides, we do not update the positions of orphan galaxies as done in \cite{Henriques2013} for simplicity, which has little impact on the results during EoR.

We then fit the free parameters based on UVLF observations from $z=6$ to 12 \cite[e.g. ][]{Bouwens2021, Adams2024ApJ, Finkelstein2024ApJ, Donnan2024MNRAS, Harikane2023ApJS}.
Due to the limit of dark matter simulation box size (i.e. $300\,{\rm Mpc}/h$) and number of tree samples, data points at the brightest end of the UVLF (see the grey area in the top panel of Fig.~\ref{fig:uv_smf_multi}) are not included in the MCMC calculations. 

\begin{table*}
\centering
\caption{List of free parameters related to the galaxy formation and evolution process in LG15 and LG20 \citep[see more details in ][]{Henriques_2015, Henriques_2020}. The parameter values showed in the column ``Fiducial'' are those originally adopted in LG20 i.e. the ones in \cite{Henriques_2020}, while those in the other columns are obtained from our MCMC fitting. In bold we highlight those which are different from the values in LG15 and LG20. Note that in the columns LG15\_bestfit and LG15\_final, the bold ones are those different to the values in \cite{Henriques_2015}.}
\begin{tabular}{lccccccc}
\hline 
Parameter & Fiducial & LG20\_bestfit & LG20\_final & LG20\_dust\_bestfit & LG20\_dust\_final & LG15\_bestfit & LG15\_final \\
\hline
\vspace{0.2em}
$\alpha_{\rm H_{2}}(\alpha_{\rm SF})$ & 0.06 & \textbf{0.11} & \textbf{0.11} & \textbf{0.19} & \textbf{0.19} & \textbf{6.1$\boldsymbol{\times 10^{-3}}$}  &  \textbf{6.1$\boldsymbol{\times 10^{-3}}$} \\
\vspace{0.2em}
$M_{\rm crit,0}$ & -- & -- & -- & -- & -- & \textbf{0.27}  &  0.24 \\
\vspace{0.2em}
$\alpha_{\rm SF,\,burst}$ & 0.50 & \textbf{0.65} & 0.5 & \textbf{0.25} & 0.5 & 0.6  &  0.6 \\
\vspace{0.2em}
$\beta_{\rm SF,\,burst}$ & 0.38 & \textbf{0.21} & \textbf{0.21} & \textbf{0.27} & \textbf{0.27} & \textbf{1.2}  & 1.9 \\
\vspace{0.2em}
$k_{\rm AGN}$ & 2.5$\times 10^{-3}$ & \textbf{1.6$\boldsymbol{\times 10^{-3}}$} & 2.5$\times 10^{-3}$ & \textbf{0.011} & 2.5$\times 10^{-3}$ & \textbf{6.2$\boldsymbol{\times 10^{-3}}$}  & 5.3$\times 10^{-3}$ \\
$f_{\rm BH}$ & 0.066 & \textbf{0.07} & 0.066 & \textbf{0.049} & 0.066 & \textbf{0.082}  & 0.041 \\
\vspace{0.2em}
$V_{\rm BH}$ & 700 & \textbf{730} & 700 & \textbf{50} & 700 & \textbf{740}  & 750 \\
\vspace{0.2em}
$\epsilon_{\rm reheat}$ & 5.6 & \textbf{1.6} & 5.6 & \textbf{0.51} & 5.6 & \textbf{1.3}  & 2.6 \\
\vspace{0.2em}
$V_{\rm reheat}$ & 110 & 110 & 110 & \textbf{150} & 110 & \textbf{320}  & 480 \\
\vspace{0.2em}
$\beta_{\rm reheat}$ & 2.9 & \textbf{4.1} & 2.9 & \textbf{3.4} & 2.9 & \textbf{0.79}  & 0.72 \\
\vspace{0.2em}
$\eta_{\rm eject}$ & 5.5 & \textbf{4.7} & 5.5 & \textbf{2.8} & 5.5 & \textbf{0.28}  & \textbf{0.28} \\
\vspace{0.2em}
$V_{\rm eject}$ & 220 & \textbf{200} & 220 & \textbf{490} & 220 & \textbf{59}  & 100 \\ 
\vspace{0.2em}
$\beta_{\rm eject}$ & 2.0 & \textbf{2.4} & 2.0 & \textbf{4.3} & 2.0 & \textbf{1.2}  & 0.8 \\
\vspace{0.2em}
$\gamma_{\rm reinc}$ & 1.2$\times 10^{10}$ & \textbf{7.7$\boldsymbol{\times 10^{9}}$} & \textbf{7.7$\boldsymbol{\times 10^{9}}$} & \textbf{9.6$\boldsymbol{\times 10^{8}}$} & \textbf{9.6$\boldsymbol{\times 10^{8}}$} & \textbf{6.5$\boldsymbol{\times 10^{10}}$}  & 3.0$\times 10^{10}$ \\
\vspace{0.2em}
$\alpha_{\rm friction}$ & 1.8 & 1.8 & 1.8 & \textbf{2.3} & 1.8 & \textbf{3.8}  & 2.5 \\
\vspace{0.2em}
$M_{\rm r.p.}$ & 4.1$\times 10^4$ & \textbf{2.3$\boldsymbol{\times 10^4}$} & \textbf{2.3$\boldsymbol{\times 10^4}$} & \textbf{2.5$\boldsymbol{\times 10^4}$} & \textbf{2.5$\boldsymbol{\times 10^4}$} & \textbf{1.4$\boldsymbol{\times 10^4}$}  & 1.2$\times 10^4$ \\ 
\hline
\end{tabular}
\label{table:LG_all_para}
\end{table*}
We perform the MCMC fitting for 15 free parameters within LG20 and 16 within LG15, which are listed in Tab.~\ref{table:LG_all_para}. 
We note that $\alpha_{\rm H_{2}}$ in LG20 is the star formation efficiency based on H$_{2}$ surface density, while $\alpha_{\rm SF}$ is the star formation efficiency adopted in LG15.
$M_{\rm crit,0}$ is the mass threshold for star formation adopted in LG15, which is not included in LG20. 
$\alpha_{\rm SF,\,burst}$ and $\beta_{\rm SF,\,burst}$ are related to the burst of star formation expected after a galactic merger.
$k_{\rm AGN}$, $f_{\rm BH}$ and $V_{\rm BH}$ are associated with the modeling of the central massive black hole growth within galaxies. 
$\epsilon_{\rm reheat}$, $V_{\rm reheat}$ and $\beta_{\rm reheat}$ regulate the gas heating through supernovae explosions, while
$\eta_{\rm eject}$, $V_{\rm eject}$ and $\beta_{\rm eject}$ model the associated gas ejection. 
$\gamma_{\rm reinc}$ is an efficiency factor which regulates the reincorporation time-scale of gas ejected in winds, 
while $\alpha_{\rm friction}$ is the efficiency factor of galaxy merger delay time due to dynamical friction. 
Finally, $M_{\rm r.p.}$ is associated with the ram-pressure stripping process. 

Considering the uncertainties of galaxy formation and dust modeling during the EoR, we perform three sets of MCMC runs, one respectively for LG20 and LG15 by assuming no dust correction to compute the UV luminosity, and one for LG20 with dust correction.
Since the UVLF after dust correction in LG15 are much lower than the observed ones \citep{Clay2015MNRAS}, we do not perform an MCMC analysis for the case with dust correction in LG15.

\section{Results} 
\label{sec:resul}
The best-fit results of free parameters in LG15 and LG20 are listed in Tab.~\ref{table:LG_all_para}, while Appendix~\ref{app:mcmc} shows some of the results of the MCMC sample analysis. 
In the column named Fiducial, we list the parameter values from \cite{Henriques_2020}, which are fitted with the observations at $z\le3$ for LG20.
The columns LG20\_bestfit and LG20\_dust\_bestfit are  the best-fit results of LG20 without and with dust correction on the UV luminosity, respectively.
The column LG15\_bestfit is the best-fit results of LG15 without dust correction. 
Since most parameters are not well constrained by UVLF observations \cite[see also the discussions in][]{Ma2023MNRAS}, in columns LG20\_final and LG20\_dust\_final we select four well limited parameters from LG20\_bestfit and LG20\_dust\_bestfit (i.e. $\alpha_{\rm H_{2}}$, $\beta_{\rm SF,\,burst}$, $\gamma_{\rm reinc}$ and $M_{\rm r.p.}$, see the Appendix~\ref{app:mcmc}), while the others are the same as those of the Fiducial model. 
Column LG15\_final includes two well-limited parameters from LG15\_bestfit ($\alpha_{\rm SF}$ and $\eta_{\rm eject}$), while the others are the original ones in LG15 that fitted low-$z$ observations \citep{Henriques_2015}. 
Note that \cite{Henriques_2015} adopted $\alpha_{\rm SF}=0.025$ and $\eta_{\rm eject}=0.62$, i.e. our best-fit $\alpha_{\rm SF}$ is only $\sim 24\%$ of the original one in LG15. 

In the following subsections, we will discuss the galaxy properties and the ionizing photon budget from LG20 and LG15 with the setting of parameter values listed in Tab.~\ref{table:LG_all_para}, i.e. the results of seven LG simulations.

\subsection{Galaxy properties}
\begin{figure*}
\centering
	\includegraphics[width=0.98\linewidth]{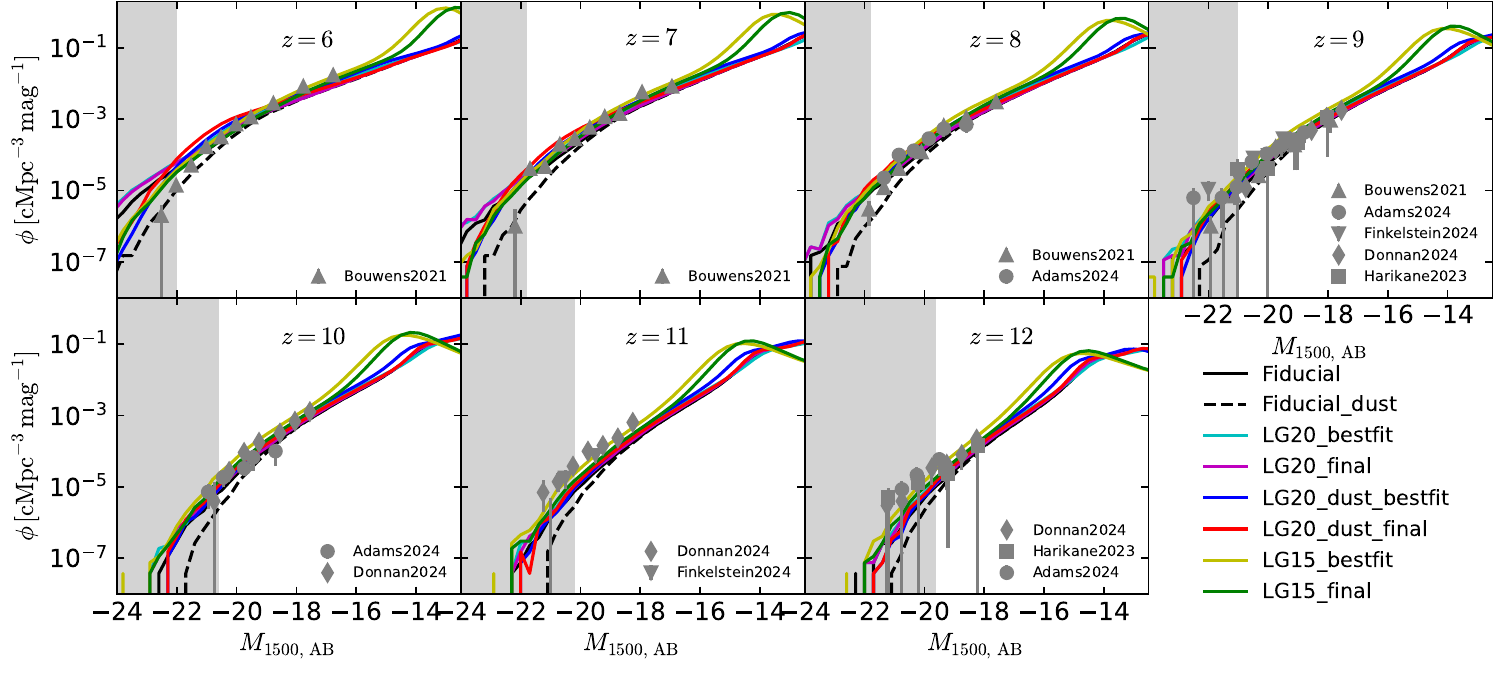}
        \includegraphics[width=0.98\linewidth]{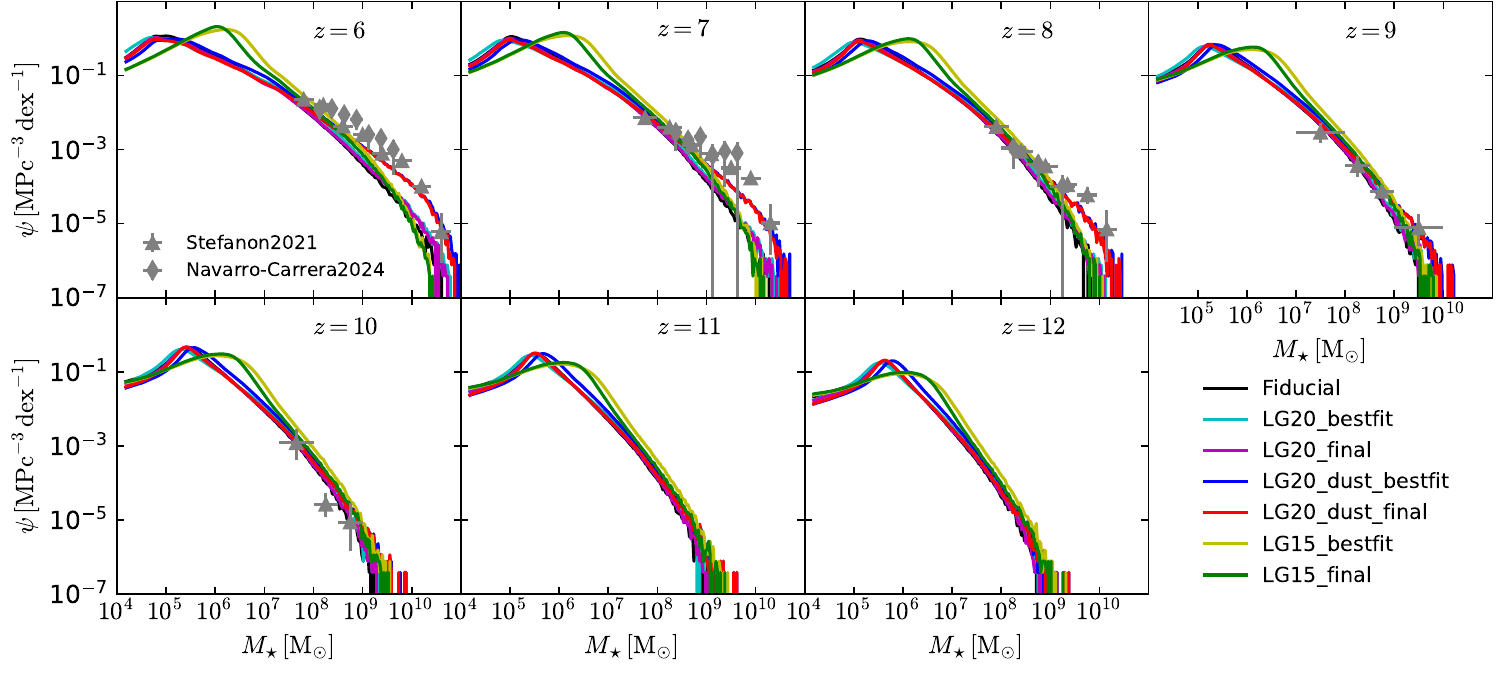}
    \caption{{\bf Top panel}: UV luminosity functions (UVLF, $\phi$) at $z=6$ to 12 from LG simulation Fiducial (black), LG20\_bestfit (cyan), LG20\_final (magenta), LG20\_dust\_bestfit (blue), LG20\_dust\_final (red), LG15\_bestfit (yellow) and LG15\_final (green).
    The dashed black line (named Fiducial\_dust) is not a new simulation, but rather Fiducial with a correction for dust. 
    The observational data points are from \citet[up triangle]{Bouwens2021}, \citet[circle]{Adams2024ApJ}, \citet[down triangle]{Finkelstein2024ApJ}, \citet[diamond]{Donnan2024MNRAS}, and \citet[square]{Harikane2023ApJS}.
    Note that the data points within the light-gray areas are not included in the MCMC calculations. 
    {\bf Bottom panel}: stellar mass function (SMF, $\psi$) from the same simulations as above. 
    The observational data points are from \citet[up triangle]{Stefanon2021} and \citet[diamond]{Navarro-Carrera2024}.
    }
    \label{fig:uv_smf_multi}
\end{figure*}
The top panel of Fig.~\ref{fig:uv_smf_multi} shows the UVLF, $\phi$, from $z=6$ to 12 from the seven LG simulations, together with observational data from HST and JWST. 
As a reference, we also show the UVLFs obtained from Fiducial with the addition of dust correction (named as Fiducial\_dust), which, as expected, are lower than those of Fiducial at the bright end (i.e. $M_{\rm 1500,\, AB}<-20$). 
Note that the results at $M_{\rm 1500,\, AB}>-16$ might not be very robust, due to the resolution limitation in the simulations.
From Fig.~\ref{fig:uv_smf_multi}, we observe that while the UVLF from Fiducial\_dust is fairly consistent with the observations of \cite{Bouwens2021} at $z=6$ and 7, they are significantly lower than JWST results at $z\ge8$.
Although the MCMC calculations do not include the data points at the bright end, the UVLF from all simulations are still consistent with observations at $z=8-10$, they are slightly higher at $M_{\rm 1500,\, AB}<-22$ and $z=6-7$, and fall below observations at $M_{\rm 1500,\, AB}<-20$ and $z=11-12$.
The UVLF from LG20\_bestfit and LG20\_final are very similar among themselves and also to  Fiducial, since the minor changes of the galaxy formation parameters do not significantly affect the UVLF \citep{Ma2023MNRAS}.
The UVLF from LG20\_dust\_bestfit and LG20\_dust\_final are consistent with Fiducial at $z\ge8$, while at $z=6$ and 7 they are slightly higher for  $-22<M_{\rm 1500,\, AB}<-19$ and become lower  at $M_{\rm 1500,\, AB}<-22$ due to dust absorption.
The UVLF from LG15\_bestfit and LG15\_final are consistent with those of Fiducial at $M_{\rm 1500\, AB}<-16$. 
The discrepancies observed at $M_{\rm 1500\, AB}>-16$ may be due to the incomplete samples from the simulations at such magnitudes, and probably also to the different prescription of star formation employed in LG15 and LG20, in particular the adoption in L15 of a critical mass threshold for star formation $M_{\rm crit,0}$, which is not present in LG20.
Since only the faint end of the UVLF is sensitive to $M_{\rm crit,0}$, this parameter in LG15 is poorly limited by current UVLF observations. 
As discussed in \cite{Henriques_2020}, the prescription adopted in LG15 allows star formation earlier than in LG20, and thus produces a larger SFR and stellar mass in  low-mass halos, resulting in a higher UVLF at the faint end. 
{ The total $\chi^2$ of UVLF from $z=6$ to 12 of the eight models shown in the top panel of Fig.~\ref{fig:uv_smf_multi} are 324.9, 283.9, 607.6, 566.5, 348.2, 1001.3, 584.2, 224.7. Note that, since some data points of UVLF observations are not included in the MCMC, these $\chi^2$ values are different to the best-fit ones of MCMC.
From the $\chi^2$ values, LG15\_final is the best in matching UVLF observations, followed by the Fiducial model with dust correction (i.e. Fiducial\_dust in Fig.~\ref{fig:uv_smf_multi}).} 

\begin{figure*}
\centering
	\includegraphics[width=0.98\linewidth]{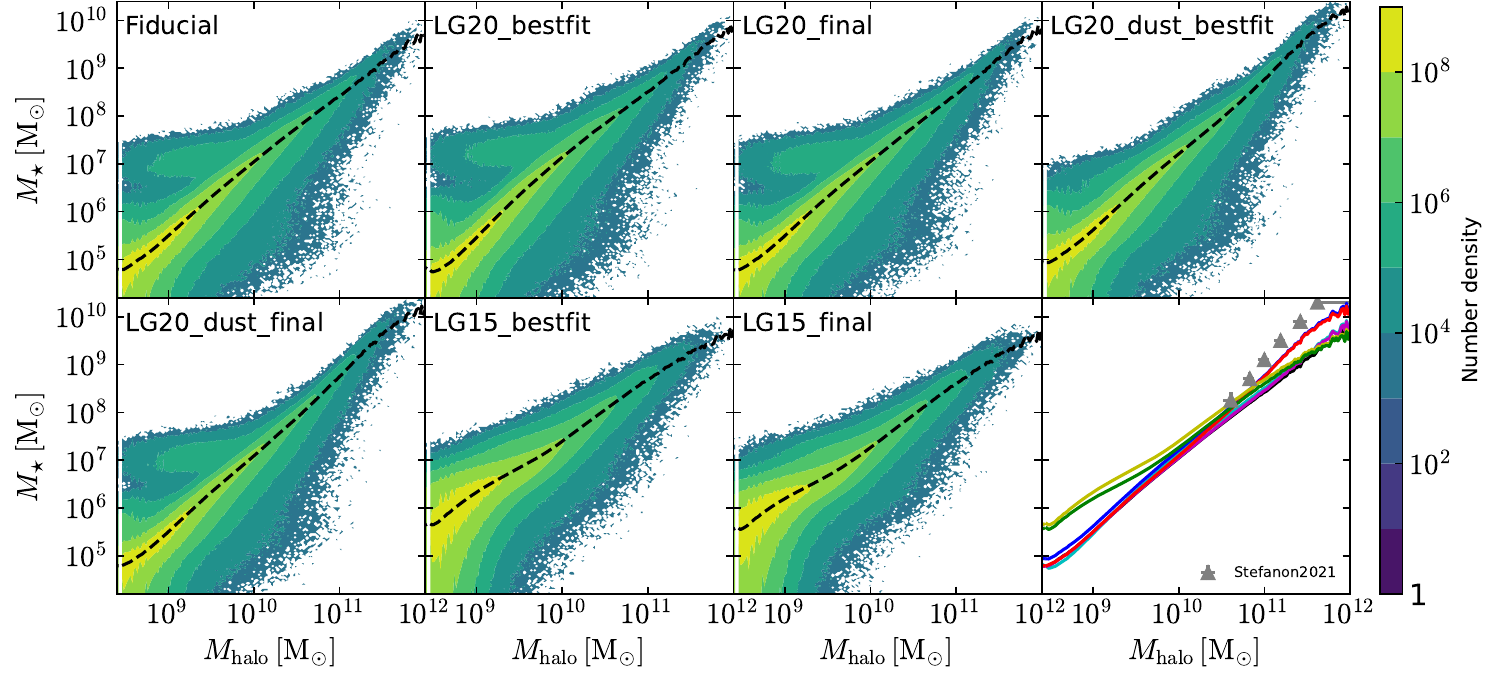}
        \includegraphics[width=0.98\linewidth]{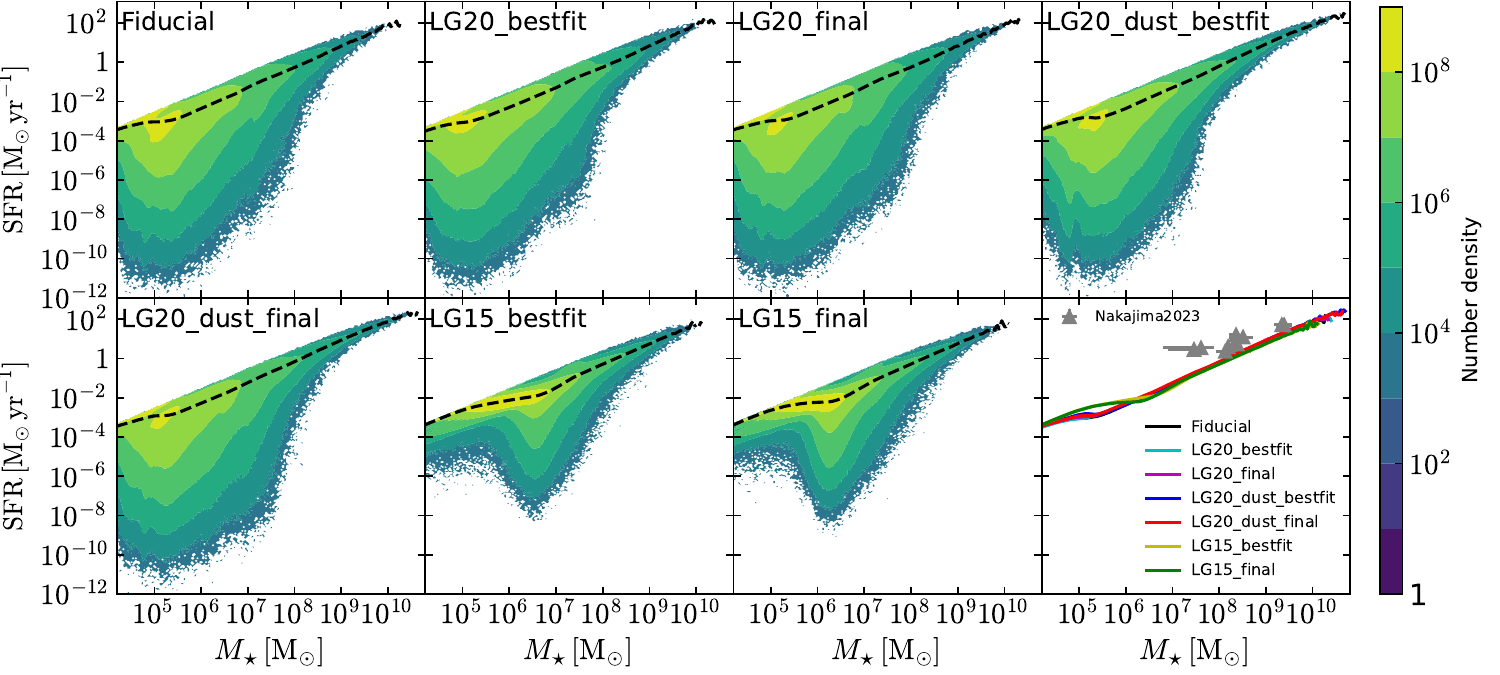}
    \caption{{\bf Top panel}: 2-D distributions of stellar mass $M_{\star}$ versus halo mass $M_{\rm halo}$ at $z=7$ from LG simulation Fiducial, LG20\_bestfit, LG20\_final, LG20\_dust\_bestfit, LG20\_dust\_final, LG15\_bestfit and LG15\_final, from left to right and from top to bottom.
    The dashed black lines are the mean $M_{\star}$ of halos with the same $M_{\rm halo}$. The mean curves from all models are also shown in the bottom-right plot with the same legend of Fig.~\ref{fig:uv_smf_multi}, together with grey observational data points at $z\approx7$ from \cite{Stefanon2021}.
    {\bf Bottom panel}: Same as above but for SFR versus $M_{\star}$.
    The grey observational data points in the bottom-right plot are those at $z=6.5-7.5$ from \cite{Nakajima2023ApJS}.
    }
    \label{fig:pro_2ddis_halo_stellar_sfr}
\end{figure*}

The bottom panel of Fig.~\ref{fig:uv_smf_multi} shows the stellar mass function (SMF, $\psi$) at $z=6$ to 12 from the seven LG simulations, together with
observational data points from \cite{Stefanon2021} and \cite{Navarro-Carrera2024}. 
We note that, due to the resolution limitation in the simulations, the results at $M_{\star}<10^{7}\,\rm M_{\odot}$ might not be very robust.
The SMF from LG20\_bestfit and LG20\_final are consistent with those from Fiducial. 
On the other hand, as shown in Tab.~\ref{table:LG_all_para}, in comparison to the Fiducial case, the inclusion of dust correction within the MCMC calculation results in a higher star formation efficiency $\alpha_{\rm H_{2}}$, and in a lower gas reincorporation time-scale factor $\gamma_{\rm reinc}$, leading to a  higher stellar mass in the massive halos (see also the top panel of Fig.~\ref{fig:pro_2ddis_halo_stellar_sfr}). As a consequence, LG20\_dust\_bestfit and LG20\_dust\_final show a SMF at $M_{\star}>10^{9}\,\rm M_{\odot}$ and $z \le 9$ higher than those of Fiducial. 
The SMF from LG15\_bestfit and LG15\_final are consistent with those of Fiducial at $M_{\star}>10^{7}\,\rm M_{\odot}$, while they differ at $M_{\star}<10^{7}\,\rm M_{\odot}$ because of the incomplete samples from the simulations at such masses and the different prescription for star formation adopted in LG15 and LG20. 
Although SMF observations are not included in the MCMC calculations, the SMF from LG20\_dust\_bestfit and LG20\_dust\_final match well with data from \cite{Stefanon2021} and \cite{Navarro-Carrera2024}, while the others are consistent with observations only at $z\ge9$, suggesting that dust correction should probably be included to fit both UVLF and SMF observations. 

The top panel of Fig.~\ref{fig:pro_2ddis_halo_stellar_sfr} shows the 2-D distributions of stellar mass $M_{\star}$ versus halo mass $M_{\rm halo}$ at $z=7$ from the seven LG simulations. 
The color-bar refers to the number density of galaxies, computed as $N_{\rm gal}/(B_{x}B_{y})$, where $B_{x}$ and $B_{y}$ are the bin-width along the $x$ and $y$-axis, and $N_{\rm gal}$ is the number of galaxies in each bin. 
We also show the mean $M_{\star}$ versus $M_{\rm halo}$, together with data at $z\approx7$ from \cite{Stefanon2021}.
The results from LG20\_bestfit and LG20\_final are very similar to those from Fiducial. 
Their $M_{\star}$ is linearly related to $M_{\rm halo}$ in the log-space, with a power-law relation $M_{\star} \approx 10^{-7.6}M_{\rm halo}^{1.46}$. 
LG20\_dust\_bestfit and LG20\_dust\_final typically have $M_{\star}$ larger than that of Fiducial at $M_{\rm halo}>10^{10.5}\,\rm M_{\odot}$, due to the higher star formation efficiency $\alpha_{\rm H_{2}}$ and lower reincorporation time-scale factor $\gamma_{\rm reinc}$.
Because of this, the index of the power-law relation between $M_{\star}$ and $M_{\rm halo}$ is increased to $\sim 1.59$. 
Note that the relation of $M_{\star}$ and $M_{\rm halo}$ from LG20\_dust\_bestfit and LG20\_dust\_final is more consistent with the observations of \cite{Stefanon2021} at $z\approx7$ than in other models.
The $M_{\star}$ of LG15\_bestfit and LG15\_final have a much larger scatter than Fiducial at $M_{\rm halo}>10^{11}\,\rm M_{\odot}$, while their mean  $M_{\star}$ are consistent with the latter.
At $M_{\rm halo}<10^{10}\,\rm M_{\odot}$, the $M_{\star}$ of LG15\_bestfit and LG15\_final are typically higher than Fiducial, as the prescription for star formation adopted in LG15 allows an earlier formation of stars in low-mass halos. This results in an index of the power-law relation between $M_{\star}$ and $M_{\rm halo}$ of $\sim 1.2$, lower than Fiducial.

The bottom panel of Fig.~\ref{fig:pro_2ddis_halo_stellar_sfr} shows the 2-D distributions of SFR versus stellar mass $M_{\star}$ at $z=7$ from the seven LG simulations, together with 
the mean SFR and observational results from galaxies at $z=6.5-7.5$ by \cite{Nakajima2023ApJS}.
The SFR distributions of LG20\_bestfit and LG20\_final are similar to those of Fiducial, with a very large scatter at $M_{\star}<10^{8}\,\rm M_{\odot}$.
The SFR can be as low as $<10^{-10}\,\rm M_{\odot}\,yr^{-1}$ for galaxies with $M_{\star} \sim 10^{5}\,\rm M_{\odot}$, due to the supernovae feedback that can efficiently quench star formation in some low-mass galaxies. 
As mentioned earlier, the LG20\_dust\_bestfit and LG20\_dust\_final have more massive galaxies than Fiducial, while their SFR versus $M_{\star}$ relations are consistent with the latter.
The galaxies from LG15\_bestfit and LG15\_final have globally higher SFR than Fiducial at $M_{\star}<10^{6}\,\rm M_{\odot}$, and especially they have no SFR lower than $10^{-5}\,\rm M_{\odot}\,yr^{-1}$ at $M_{\star}<10^{5.5}\,\rm M_{\odot}$.
The discrepancies observed in the SFR distributions of LG15 and LG20 are due again to the incomplete samples from the simulations in such mass range and to the  different prescriptions adopted for star formation. 
We note that our estimated SFR versus $M_{\star}$ relations in all simulations are roughly consistent with observations at $z=6.5-7.5$ from \cite{Nakajima2023ApJS}.

\begin{figure}
\centering
        \includegraphics[width=1.0\linewidth]{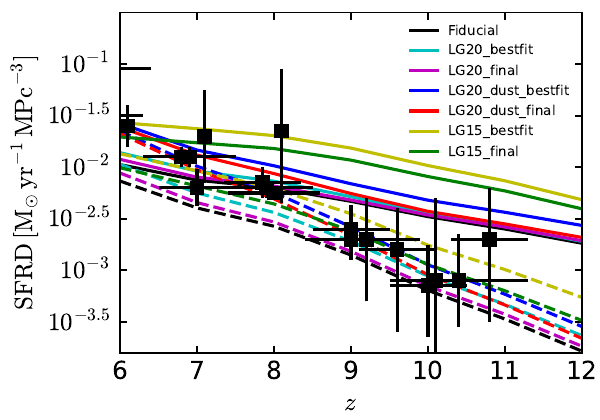}
    \caption{Redshift evolution of the star formation rate density (SFRD) from LG simulation Fiducial (black), LG20\_bestfit (cyan), LG20\_final (magenta), LG20\_dust\_bestfit (blue), LG20\_dust\_final (red), LG15\_bestfit (yellow) and LG15\_final (green).
    The solid lines refer to results when all galaxies are included, while the dashed lines include only galaxies with $M_{\star} \ge 10^{7}\,\rm M_{\odot}$.
    The observational data points are those summarized in \cite{Ma2017MNRAS}.
    }
    \label{fig:sfrd}
\end{figure}
Fig.~\ref{fig:sfrd} shows two cases of volume averaged SFR density (SFRD) as a function of $z$,
one including all galaxies with non-zero star formation, while the other one considers only those with $M_{\star} \ge 10^{7}\,\rm M_{\odot}$ (as suggested from the bottom panel of Fig.~\ref{fig:uv_smf_multi} and the top panel of Fig.~\ref{fig:nion_dis_vz}).
As a comparison, we also show the observational SFRD summarized in \cite{Ma2017MNRAS}.
When all galaxies are included, the SFRD from G20\_bestfit and LG20\_final are similar to those from Fiducial, while
LG20\_dust\_bestfit and LG20\_dust\_final produce higher SFRD, especially at $z\le8$, due to their larger star formation efficiency $\alpha_{\rm H_{2}}$.
LG15\_bestfit and LG15\_final have the highest SFRD, which is $\sim 0.5$ dex larger than in Fiducial. 
When galaxies with $M_{\star} < 10^{7}\,\rm M_{\odot}$ are removed, the SFRD becomes smaller. For example, the one of Fiducial is reduced by $\sim 30\%$ and $60\%$ at $z=6$ and 12, respectively.
The SFRD of LG15\_bestfit and LG15\_final are reduced by $\sim 50\%$ at $z=6$, as LG15 predicts more galaxies at $M_{\star} < 10^{7}\,\rm M_{\odot}$ than Fiducial (see Fig.~\ref{fig:uv_smf_multi}).
Since LG20\_dust\_bestfit and LG20\_dust\_final have more massive galaxies than Fiducial, their SFRD are only reduced by $\sim 15\%$ and become the highest at $z=6$. 
Due to the large error-bars of the data points, the SFRD from all simulations are roughly consistent with observations. 
Note that, since the data points do not include faint galaxies \cite[e.g. ][]{Bouwens2015ApJ}, the SFRD of galaxies with $M_{\star} \ge 10^{7}\,\rm M_{\odot}$ are more consistent with observations at $z>9$ than those including all galaxies.

\begin{figure*}
\centering
	\includegraphics[width=0.98\linewidth]{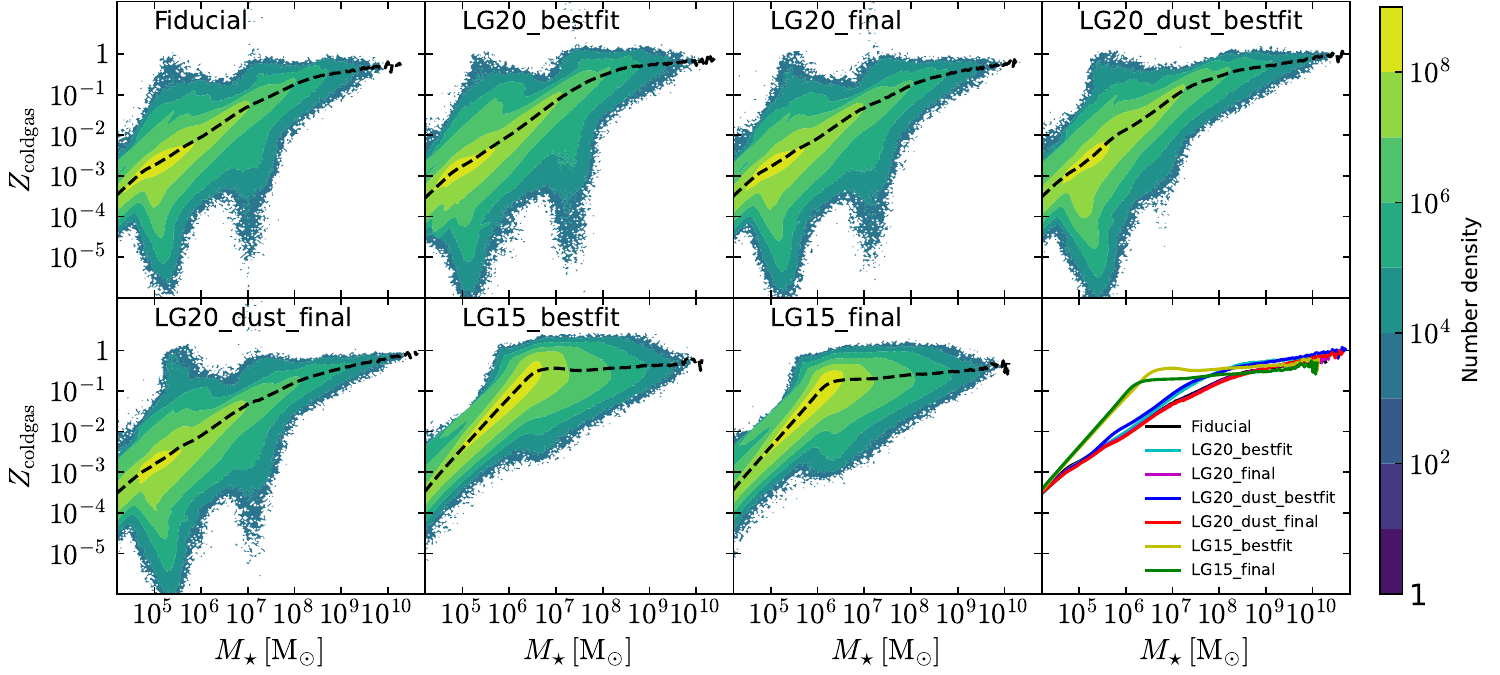}
        \includegraphics[width=0.98\linewidth]{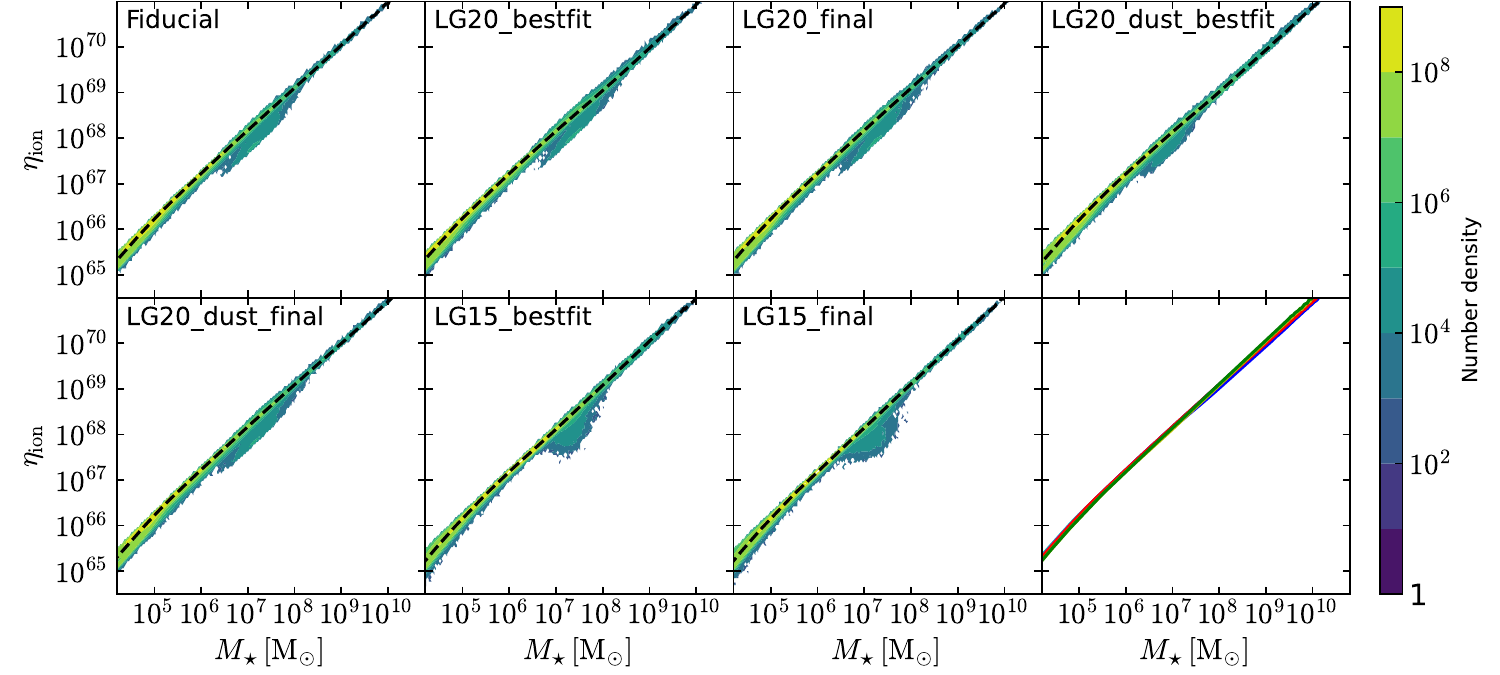}
    \caption{{\bf Top panel}: 2-D distributions of cold gas metallicity $Z_{\rm coldgas}$ versus stellar mass $M_{\star}$ of galaxies at $z=7$ from LG simulation Fiducial, LG20\_bestfit, LG20\_final, LG20\_dust\_bestfit, LG20\_dust\_final, LG15\_bestfit and LG15\_final, from left to right and from top to bottom.
    The dashed black lines are the mean $Z_{\rm coldgas}$ of galaxies with the same $M_{\star}$. The mean curves from all models are also shown in the bottom-right plot with the same legend of Fig.~\ref{fig:uv_smf_multi}. 
    {\bf Bottom panel}: Same as above but for the time integrated ionizing photon number $\eta_{\rm ion}$ versus stellar mass $M_{\star}$.
    }
    \label{fig:pro_2ddis_stellar_metal_ionint}
\end{figure*}
The top panel of Fig.~\ref{fig:pro_2ddis_stellar_metal_ionint} shows the 2-D distributions of cold gas metallicity $Z_{\rm coldgas}$ as a function of stellar mass $M_{\star}$ of galaxies at $z=7$ from all LG simulations, where  $Z_{\rm coldgas} = M_{\rm metal}/M_{\rm coldgas}/0.02$, with $M_{\rm coldgas}$ galactic cold gas mass, $M_{\rm metal}$ metal mass within cold gas, and $0.02$ is the solar metallicity. 
Since the dust absorption of UV photons relates to the metallicity of cold gas, $Z_{\rm coldgas}$ is important to compute the UVLF when considering the dust correction. 
LG20\_bestfit and LG20\_final show $Z_{\rm coldgas}$ distributions similar to those of Fiducial.
Their mean $Z_{\rm coldgas}$ increases linearly in log-space with increasing $M_{\star}$  at $M_{\star}<10^{8}\,\rm M_{\odot}$, while the increase is less rapid for $M_{\star}>10^{8}\,\rm M_{\odot}$.
Although LG20\_dust\_bestfit and LG20\_dust\_final have more massive galaxies than Fiducial, their $Z_{\rm coldgas}$ relation with $M_{\star}$ is similar to that of Fiducial. 
The scatter of $Z_{\rm coldgas}$ from LG15\_bestfit and LG15\_final is much smaller than in Fiducial at $M_{\star}<10^{6}\,\rm M_{\odot}$, but it becomes larger at $M_{\star}>10^{8}\,\rm M_{\odot}$. 
The mean $Z_{\rm coldgas}$ from LG15\_bestfit and LG15\_final is much higher than the one from LG20 at $10^{5}\,\rm M_{\odot}<M_{\star}<10^{8}\,\rm M_{\odot}$, due to the lack of modelling time delay effects of the supernovae explosion which enrich the gas in LG15.
Such differences can explain why dust correction clearly affects the UVLF shown in \cite{Clay2015MNRAS} and computed with LG15, but has a small effect in the LG20 simulations (see the top panel of Fig.~\ref{fig:uv_smf_multi}).

\subsection{Budget of ionizing photons}
\begin{figure*}
\centering
        \includegraphics[width=0.98\linewidth]{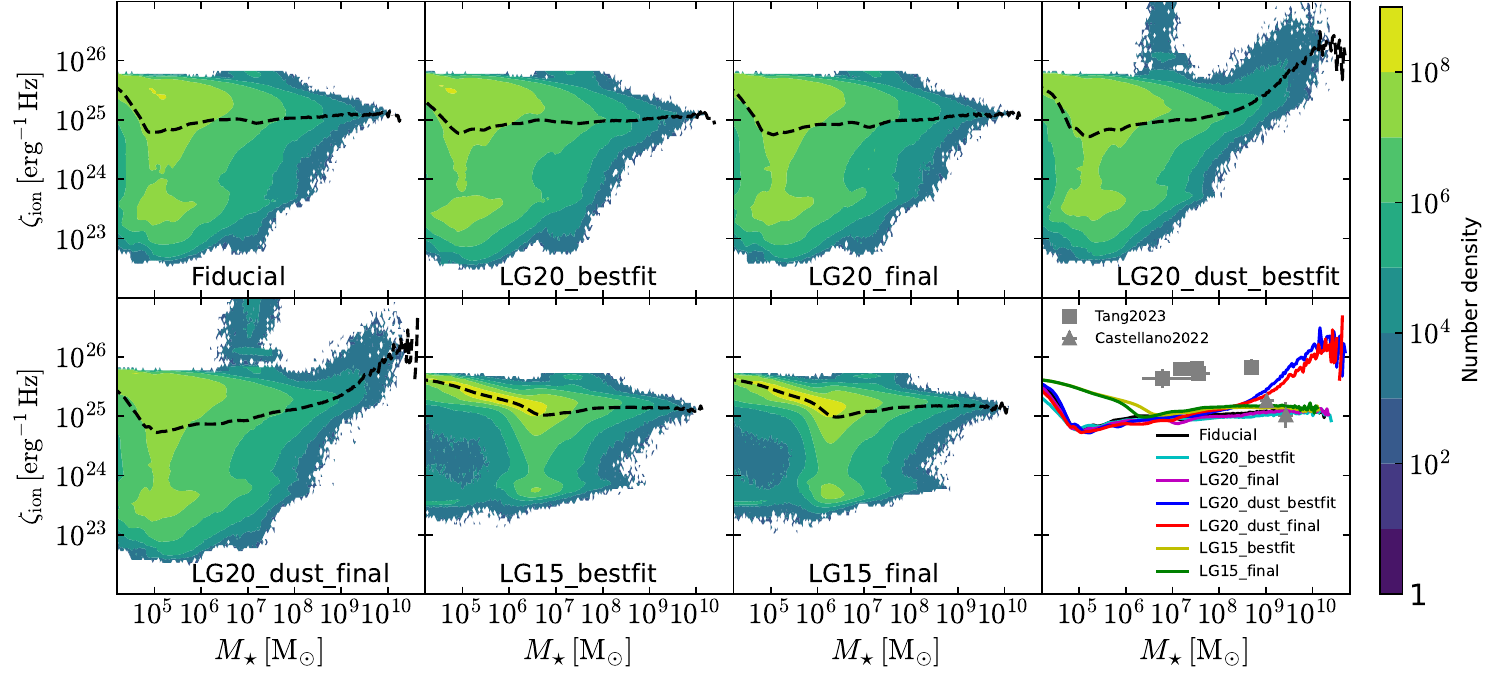}
	\includegraphics[width=0.98\linewidth]{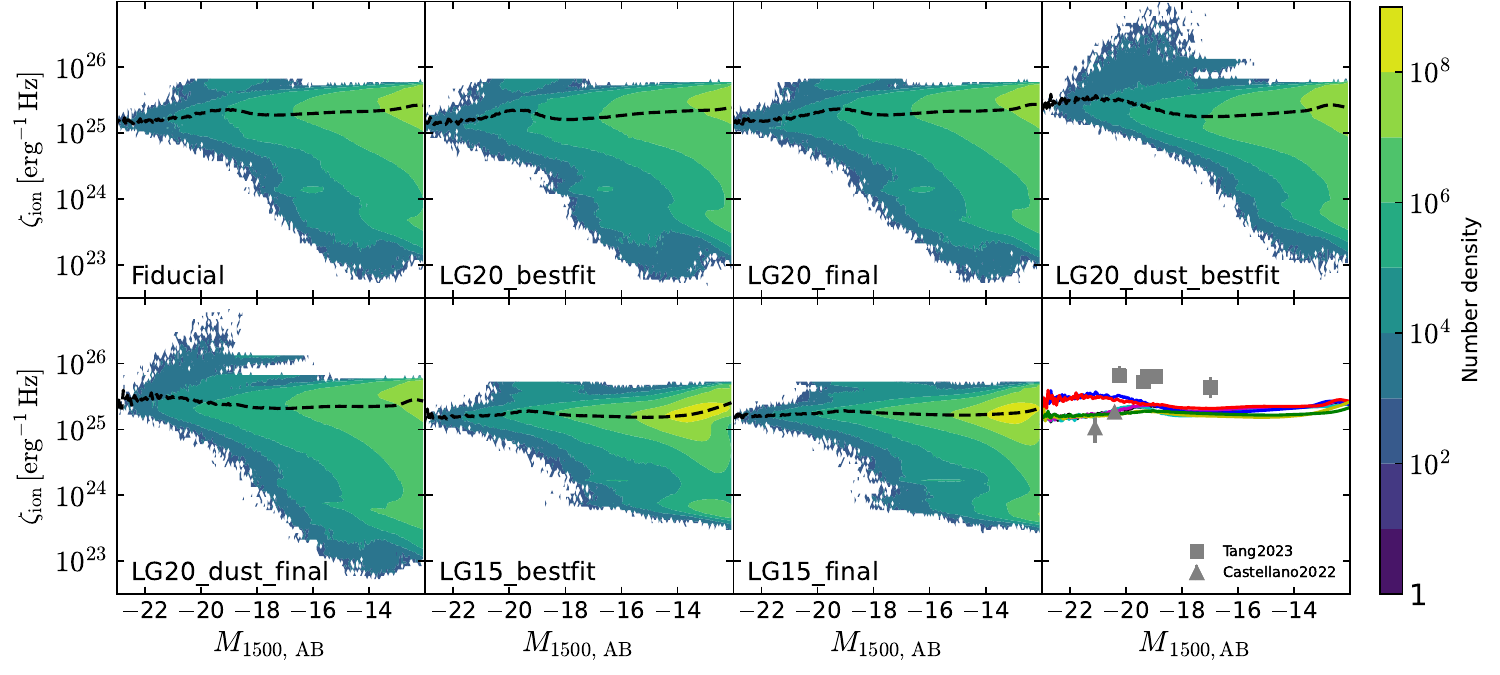}
    \caption{{\bf Top panel}: 2-D distributions of ionizing photon emission efficiency $\zeta_{\rm ion}$ versus stellar mass $M_{\star}$ at $z=7$ from LG simulation Fiducial, LG20\_bestfit, LG20\_final, LG20\_dust\_bestfit, LG20\_dust\_final, LG15\_bestfit and LG15\_final, from left to right and from top to bottom.
    The dashed black lines are the mean $\zeta_{\rm ion}$ of galaxies with the same $M_{\star}$. The mean curves from all models are also shown in the bottom-right plot with the same legend of Fig.~\ref{fig:uv_smf_multi}, together with observational data points from \citet[][square]{Tang2023} and \citet[][triangle]{Castellano2022}.
    {\bf Bottom panel}: Same as above but for $\zeta_{\rm ion}$ versus $M_{\rm 1500\,AB}$.
    }
\label{fig:pro_2ddis_stellar_mag_ioneff}
\end{figure*}

The bottom panel of Fig.~\ref{fig:pro_2ddis_stellar_metal_ionint} shows the 2-D distributions of time integrated ionizing photon number $\eta_{\rm ion}$ versus stellar mass $M_{\star}$ at $z=7$ for all LG simulations. 
Following \cite{Ma2023MNRAS} and \cite{Liu2024ApJ}, $\eta_{\rm ion}$ is computed from the star formation history and metal enrichment of galaxies obtained from the LG simulations, combined with the SPS SEDs from BPASS, and it is the total ionizing photon number emitted from the birth of galaxies until the redshift under consideration (i.e. $z=7$ in the bottom panel of Fig.~\ref{fig:pro_2ddis_stellar_metal_ionint}). 
Although the simulations show some clear differences in the stellar mass, SFR, UV luminosity $M_{\rm 1500\,AB}$ and ionizing photon emission efficiency $\zeta_{\rm ion}$ (see Fig.~\ref{fig:pro_2ddis_stellar_mag_ioneff} later on), their $\eta_{\rm ion}$ are similar and exhibit a very small scatter at the same $M_{\star}$. This is because the integration along the star formation history of galaxies reduces the scatter caused e.g. by star bursts, supernovae and AGN feedback, and other processes associated to galaxy formation. 
In all LG simulations, $\eta_{\rm ion}$ is linearly related to  $M_{\star}$ in the log-space:
\begin{equation}
\label{eq:eta_fit}
    \eta_{\rm ion} = A_{\rm ion} \left(\frac{M_{\star}}{10^{10}\,\rm M_{\odot}}\right)^{\alpha_{\rm ion}},
\end{equation}
where $A_{\rm ion}$ is the efficiency factor and $\alpha_{\rm ion}$ is the power-law index.

\begin{figure}
\centering
	\includegraphics[width=1.0\linewidth]{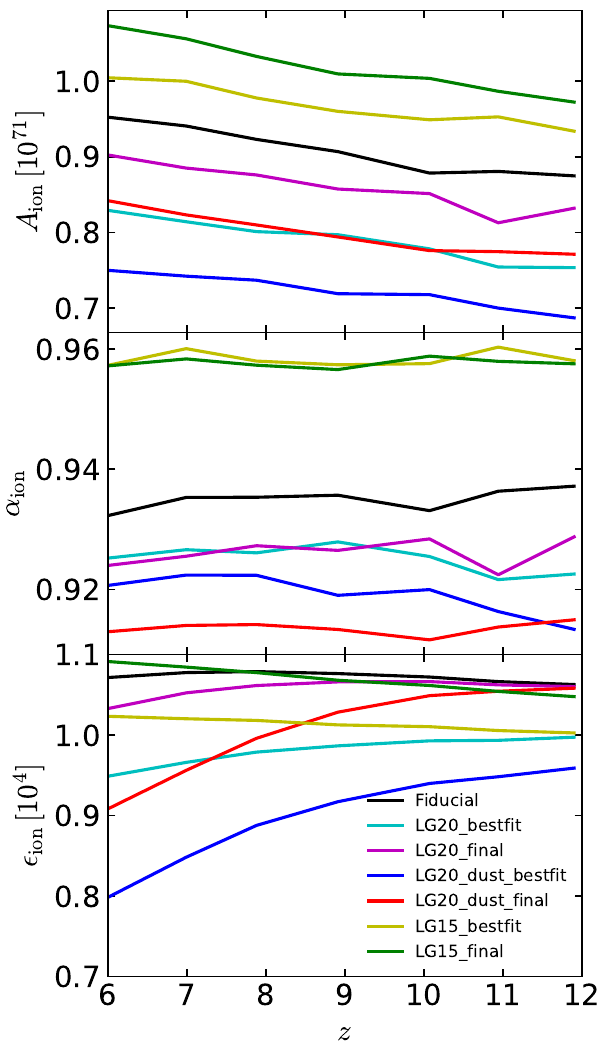}
    \caption{{\bf Top panel}: redshift evolution of the efficiency factor $A_{\rm ion}$ in Eq.~\ref{eq:eta_fit} for LG simulation Fiducial (black), LG20\_bestfit (cyan), LG20\_final (magenta), LG20\_dust\_bestfit (blue), LG20\_dust\_final (red), LG15\_bestfit (yellow) and LG15\_final (green). 
    {\bf Central panel}: as the top panel, but for the power-law index $\alpha_{\rm ion}$ in Eq.~\ref{eq:eta_fit}.
    {\bf Bottom panel}: as the top panel, but for the ionizing photon emitted per baryon $\epsilon_{\rm ion}$. 
    }
    \label{fig:ion_fittings}
\end{figure}
The evolution of $A_{\rm ion}$ and $\alpha_{\rm ion}$ from $z=12$ to 6 is shown in the top and central panels of Fig.~\ref{fig:ion_fittings}, respectively.
 We note that the above quantities are computed including only galaxies with $M_{\star}>10^{7}\,\rm M_{\odot}$, because of the incomplete sample of galaxies with lower $M_{\star}$.
$A_{\rm ion}$ for all simulations is $\sim 10^{71}$, and it increases slightly with decreasing $z$ ($\sim 10\%$ from $z=12$ to 6).
$A_{\rm ion}$ of LG20\_bestfit, LG20\_final, LG20\_dust\_bestfit and LG20\_dust\_final is $\sim 15\%$, $\sim 5\%$, $\sim 23\%$ and $\sim 13\%$ lower than the one from Fiducial, respectively.
On the other hand, $A_{\rm ion}$ of LG15\_bestfit and LG15\_final is $\sim 5\%$ and $\sim 12\%$ higher than in Fiducial, respectively.
Differently, $\alpha_{\rm ion}$ from all simulations hardly displays a redshift evolution, with values in the range [0.9-1], which denotes a good linear relation between $\eta_{\rm ion}$ and $M_{\star}$.
The $\alpha_{\rm ion}$ of LG20\_bestfit and LG20\_final are slightly lower than Fiducial, while the ones of LG20\_dust\_bestfit and LG20\_dust\_final are much lower.
The LG15\_bestfit and LG15\_final have $\alpha_{\rm ion}$ higher than Fiducial.

We also estimate the ionizing photon number emitted per baryon $\epsilon_{\rm ion}$ within stars hosted in galaxies with $M_{\star}>10^{7}\,\rm M_{\odot}$, which is defined as:
\begin{equation}
    \epsilon_{\rm ion} = \frac{\int \eta_{\rm ion} \psi {\rm d}\,M_{\star}}{({\rm M_{\odot}}/m_{\rm p})\int M_{\star} \psi {\rm d}\,M_{\star}}
\end{equation}
where $\psi$ is the SMF of galaxies, and $m_{\rm p}$ is the proton mass. 
The redshift evolution of  $\epsilon_{\rm ion}$ is displayed in the bottom of Fig.~\ref{fig:ion_fittings}.
The $\epsilon_{\rm ion}$ of Fiducial is $\sim 1.1 \times 10^4$ at $z=12$, and shows a very weak redshift evolution.
The $\epsilon_{\rm ion}$ of LG20\_final is similar to the one of Fiducial, while the one of LG20\_bestfit is $\sim 1.0\times  10^4$ at $z=12$ and reduces to $\sim 0.94\times  10^4$ at $z=6$.
The $\epsilon_{\rm ion}$ of LG20\_dust\_final is similar to Fiducial at $z=12$ but decreases to $\sim 0.91 \times 10^4$ at $z=6$.
The one of LG20\_dust\_bestfit is $\sim 0.96\times 10^4$ at $z=12$, but reduces quickly with decreasing $z$, and it is only $\sim 0.8\times  10^4$ at $z=6$.
Differently to Fiducial, the $\epsilon_{\rm ion}$ of LG15\_bestfit and LG15\_final slightly increases with decreasing $z$, becoming $\sim 1.02\times  10^4$ and $\sim 1.1\times 10^4$ at $z=6$, respectively.

\begin{figure}
\centering
        \includegraphics[width=1.0\linewidth]{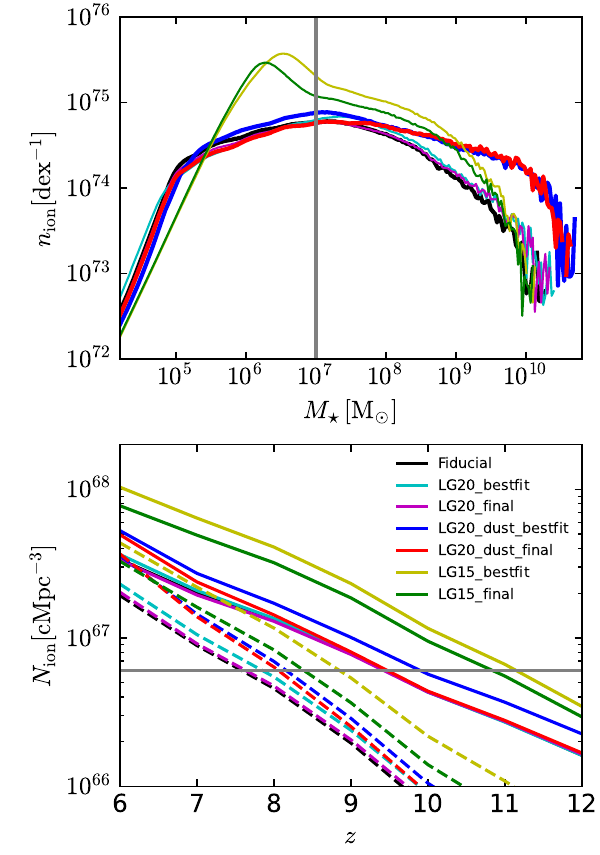}
    \caption{{\bf Top panel}: distribution of ionizing photon number $n_{\rm ion}$ as a function of $M_{\star}$ at $z=7$ from LG simulation Fiducial (black), LG20\_bestfit (cyan), LG20\_final (magenta), LG20\_dust\_bestfit (blue), LG20\_dust\_final (red), LG15\_bestfit (yellow) and LG15\_final (green). 
    The gray vertical line denotes the position of $M_{\star}=10^7\,\rm M_{\odot}$.
    {\bf Bottom panel}: evolution of ionizing photon number density $N_{\rm ion}$ from the same LG simulations as above. 
    The solid lines refer to the $N_{\rm ion}$ of all galaxies, while the dashed lines are calculated including only galaxies with $M_{\star} \ge 10^{7}\,\rm M_{\odot}$.
    The horizontal gray line denotes the total Hydrogen atom number density $N_{\rm H}$ in the Universe. 
    }
    \label{fig:nion_dis_vz}
\end{figure}
The top panel of Fig.~\ref{fig:nion_dis_vz} shows the distributions of ionizing photon number $n_{\rm ion}$ as a function of $M_{\star}$ at $z=7$ for all LG simulations, where the ionizing photon number density in the $i$th mass bin is defined as $n_{\rm ion}^{i} =\sum^{i} \eta_{\rm ion} / \Delta {\rm log}_{10} M_{\star}$, with $\Delta {\rm log}_{10} M_{\star} = 0.03$ bin-width and $\sum^{i}$ sum of $\eta_{\rm ion}$ of all galaxies within the $i$th bin.
LG20\_bestfit and LG20\_final have a $n_{\rm ion}$ distribution similar to that of Fiducial, while  LG20\_dust\_bestfit and LG20\_dust\_final have a higher distribution at $M_{\star}>10^{8}\, {\rm M}_{\odot}$, due to their larger SMF (see Fig.~\ref{fig:uv_smf_multi}).
The $n_{\rm ion}$ of LG15\_bestfit and LG15\_final are typically in between those of Fiducial and of LG20\_dust\_bestfit/LG20\_dust\_final at $M_{\star}>10^{9}\,{\rm M}_{\odot}$.

The bottom panel of Fig.~\ref{fig:nion_dis_vz} shows the volume averaged time integrated ionizing photon number densities $N_{\rm ion}$ estimated from all LG simulations, where $N_{\rm ion} = \sum \eta_{\rm ion} / V_{\rm box}$,
with $\sum$ denoting a sum over the galaxies, and $V_{\rm box}$ the volume of the simulation box. 
In one case (solid lines)  $N_{\rm ion}$ is calculated including all galaxies, while in another one (dashed) only those with $M_{\star} \ge 10^{7}\,\rm M_{\odot}$ are considered.
As a reference, we also show the total number density of Hydrogen atom $N_{\rm H}$ in the Universe, including both the neutral and ionized ones. 
For an escape fraction of ionizing photons of $100\%$, $N_{\rm H}$ would roughly denote the number density of ionizing photons needed to fully ionize the Universe, although it should be noted that gas recombination can substantially increase the required number of photons, in particular in high density regions.
When all galaxies are included, the $N_{\rm ion}$ produced in our Fiducial case is above $N_{\rm H}$ at $z<9.5$, and it is $\sim 5$ times $N_{\rm H}$ at $z=6$. 
LG20\_bestfit and LG20\_final produce similar results.
With LG20\_dust\_bestfit and LG20\_dust\_final we obtain $\sim 50\%$ more ionizing photons than Fiducial at $z=6$, while  LG15\_bestfit and LG15\_final produce $\sim 200\%$ and $\sim 100\%$ more than Fiducial, respectively.
When including only galaxies with $M_{\star} \ge 10^{7}\,\rm M_{\odot}$, the estimated $N_{\rm ion}$ of Fiducial remains above $N_{\rm H}$ at $z<7.5$, and it is $\sim 3$ times $N_{\rm H}$ at $z=6$. 
The LG20\_dust\_bestfit and LG20\_dust\_final simulations have $\sim 70\%$ more ionizing photons than Fiducial at the same redshift, while LG15\_bestfit and LG15\_final have $> 100\%$ and $\sim 70\%$ larger $N_{\rm ion}$ than Fiducial, respectively.
We note that with the assumption of an escape fraction of $10\%$, only LG15\_bestfit and LG15\_final including all galaxies would have $0.1 \times N_{\rm ion}>N_{\rm H}$ at $z = 6$, while most cases have $0.2 \times N_{\rm ion}>N_{\rm H}$ at $z = 6$, i.e. an escape fraction of $20\%$ would be needed to produce a number of photons equal to the number of Hydrogen atoms.

The top panel of Fig.~\ref{fig:pro_2ddis_stellar_mag_ioneff} shows the 2-D distributions of ionizing photon emission efficiency $\zeta_{\rm ion}$ versus stellar mass $M_{\star}$ at $z=7$ from all LG simulations, where $\zeta_{\rm ion}$  is defined as the ionizing photon emitted per second divided by the UV luminosity (i.e. unit $\rm erg\,s^{-1}\,Hz^{-1}$) at $\lambda = 1500$\AA \cite[see e.g. ][]{Wilkins2016MNRAS, Liu2024ApJ}.
As a comparison, we also show the observational results from \cite{Castellano2022} and \cite{Tang2023}.
The LG20\_bestfit and LG20\_final have  $\zeta_{\rm ion}$ distributions similar to those of Fiducial, with a very large scatter at $M_{\star}<10^{8}\,\rm M_{\odot}$, spanning the range $\zeta_{\rm ion} = 10^{(22.6-25.8)} \rm erg^{-1} \, Hz$. 
Galaxies in LG20\_dust\_bestfit and LG20\_dust\_final have a $\zeta_{\rm ion}$ at $M_{\star}>10^{8}\,\rm M_{\odot}$ higher than Fiducial, as dust absorption reduces the UV luminosities of massive galaxies, leading to a higher $\zeta_{\rm ion}$.
Some galaxies in LG20\_dust\_bestfit and LG20\_dust\_final can even have $\zeta_{\rm ion}>10^{26} \rm erg^{-1} \, Hz$ at $M_{\star}>10^{9}\,\rm M_{\odot}$.
LG15\_bestfit and LG15\_final have $\zeta_{\rm ion}$ higher than Fiducial at $M_{\star}<10^{6}\,\rm M_{\odot}$, due to their larger SFR, as shown in the bottom of Fig.~\ref{fig:pro_2ddis_halo_stellar_sfr}.
Our estimated $\zeta_{\rm ion}$ is consistent with observations from \cite{Castellano2022}, while they are lower than the ones from \cite{Tang2023}.

The bottom panel of Fig.~\ref{fig:pro_2ddis_stellar_mag_ioneff} shows the 2-D distributions of $\zeta_{\rm ion}$ versus $M_{\rm 1500\,AB}$ at $z=7$ from all LG simulations, together with observational results from \cite{Castellano2022} and \cite{Tang2023}. 
As in the top panel of the figure, LG20\_bestfit and LG20\_final have $\zeta_{\rm ion}$ distributions similar to those of Fiducial.
As dust correction reduces the UV luminosity of bright galaxies, the $\zeta_{\rm ion}$ of LG20\_dust\_bestfit and LG20\_dust\_final are higher than in Fiducial at $M_{\rm 1500\,AB}<-20$.
The $\zeta_{\rm ion}$ of LG15\_bestfit and LG15\_final show a smaller scatter than Fiducial, while their mean $\zeta_{\rm ion}$ versus $M_{\rm 1500\,AB}$ are consistent with the latter. 
Similarly to the top panel of the figure, our estimated $\zeta_{\rm ion}$ is consistent with the observational results from \cite{Castellano2022}, while globally lower than the ones of \cite{Tang2023}.

{
\begin{figure}
\centering
        \includegraphics[width=1.0\linewidth]{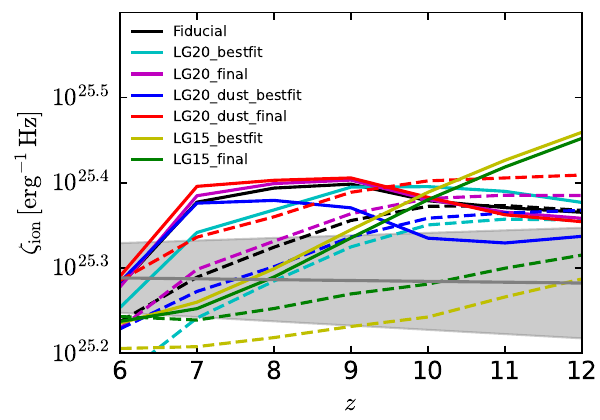}
    \caption{Redshift evolution of the average $\zeta_{\rm ion}$ from LG simulation Fiducial (black), LG20\_bestfit (cyan), LG20\_final (magenta), LG20\_dust\_bestfit (blue), LG20\_dust\_final (red), LG15\_bestfit (yellow) and LG15\_final (green).
    The solid lines refer to results when all galaxies are included, while the dashed lines include only galaxies with $M_{1500,\,\rm AB} \le -16$. 
    The gray line refers to the fitting equation of  \citealt{Simmonds2024MNRAS} with $1-\sigma$ error-bars (shaded area).
    }
    \label{fig:zeta_evo}
\end{figure}
Fig.~\ref{fig:zeta_evo} shows the evolution of the average $\zeta_{\rm ion}$ of galaxies from the seven LG simulations as functions of $z$.
In one case (solid lines)  $\zeta_{\rm ion}$ is computed including all galaxies, while in another one (dashed lines) only those with $M_{1500,\,\rm AB} \le -16$ are considered.
As a reference, we also show the fit obtained for the measured $\zeta_{\rm ion}$ in the photometric sample of \cite{Simmonds2024MNRAS}.
When all galaxies are included, the average $\zeta_{\rm ion}$ from LG20 simulations is within $10^{25.28} - 10^{25.41} \,\rm erg^{-1}\, Hz$, which increases slightly with decreasing $z$ from $z=12$ to 9, while it decreases from $z=7$ to 6. 
The $\zeta_{\rm ion}$ from LG15 simulations shows a visible decline with decreasing $z$, from $\zeta_{\rm ion}=10^{25.45} \,\rm erg^{-1}\, Hz$ at $z=12$ to $10^{25.24} \,\rm erg^{-1}\, Hz$ at $z=6$. 
When removing the very faint galaxies, i.e. including only galaxies with $M_{1500,\,\rm AB} \le -16$, $\zeta_{\rm ion}$ of all simulations roughly decreases with decreasing $z$.
The ones of LG15 simulations are globally lower than those of LG20 simulations.
Our results are roughly consistent with the fitting equation of photometric samples from \cite{Simmonds2024MNRAS} at $z=6$, while higher than the latter at $z>9$.
Note that the fitting equation of \cite{Simmonds2024MNRAS} is the result of observations at $z=3-9$, and that at $z>9$ the curve is just an extrapolation.  
}

{
\begin{figure*}
\centering
        \includegraphics[width=0.98\linewidth]{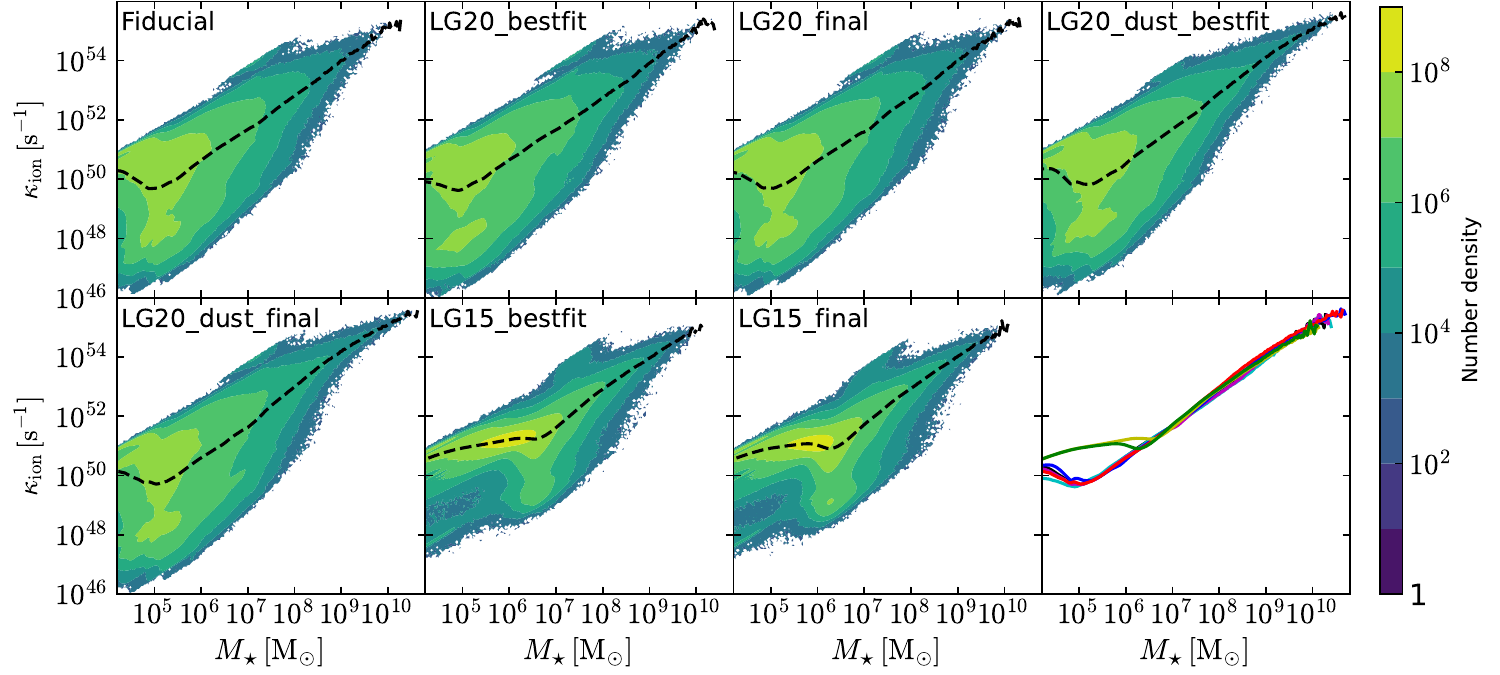}
	\includegraphics[width=0.98\linewidth]{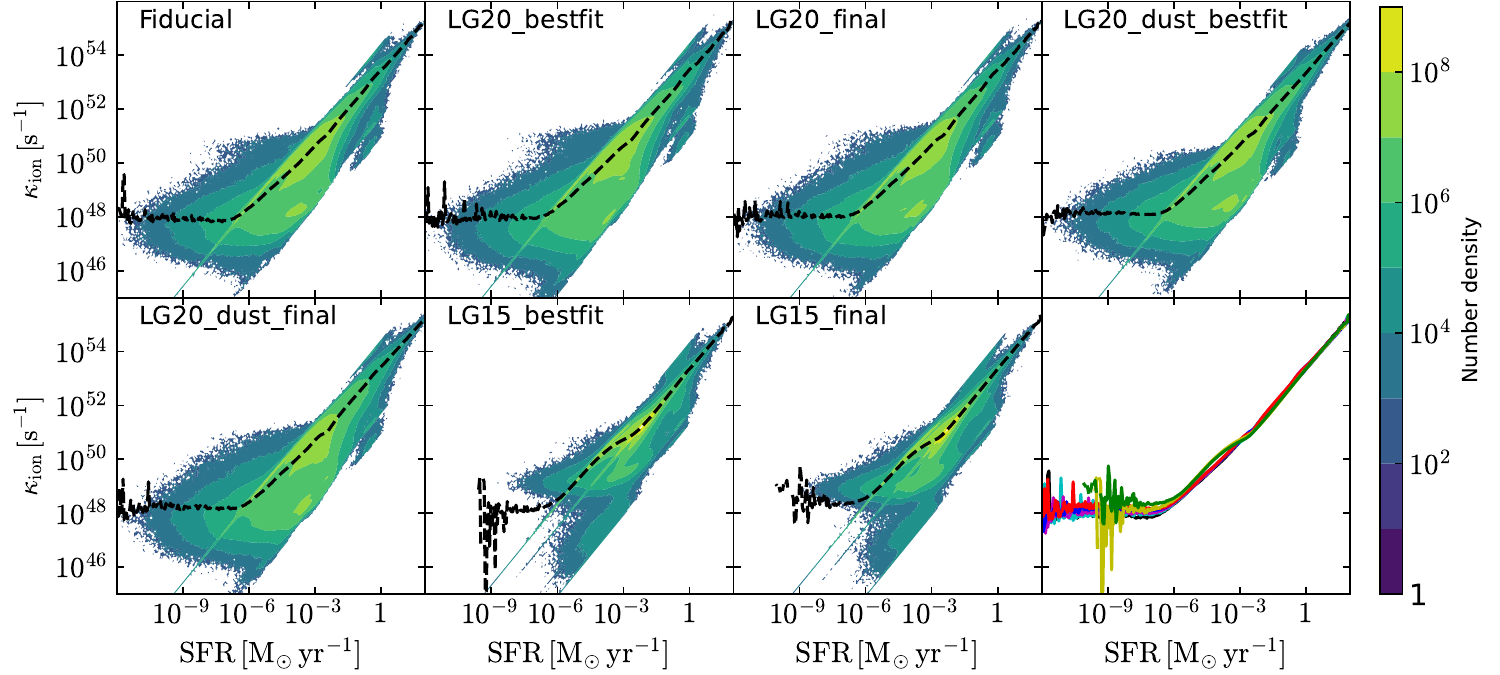}
    \caption{{\bf Top panel}: 2-D distributions of ionizing photon production efficiency $\kappa_{\rm ion}$ versus stellar mass $M_{\star}$ at $z=7$ from LG simulation Fiducial, LG20\_bestfit, LG20\_final, LG20\_dust\_bestfit, LG20\_dust\_final, LG15\_bestfit and LG15\_final, from left to right and from top to bottom.
    The dashed black lines are the mean $\kappa_{\rm ion}$ of galaxies with the same $M_{\star}$. 
    The mean curves from all models are also shown in the bottom-right plot with the same legend of Fig.~\ref{fig:uv_smf_multi}.
    {\bf Bottom panel}: Same as above but for $\kappa_{\rm ion}$ versus SFR.
    }
    \label{fig:pro_2ddis_stellar_sfr_ionflux}
\end{figure*}
The top panel of Fig.~\ref{fig:pro_2ddis_stellar_sfr_ionflux} shows the 2-D distributions of ionizing photon production efficiency $\kappa_{\rm ion}$ versus stellar mass $M_{\star}$ at $z=7$ from all LG simulations, where $\kappa_{\rm ion}$  is the ionizing photon emitted per second by galaxies.
The LG20\_bestfit, LG20\_final, LG20\_dust\_bestfit and LG20\_dust\_final have $\kappa_{\rm ion}$ distributions similar to those of Fiducial, with a large scatter at $M_{\star}<10^{8}\,\rm M_{\odot}$. 
LG15\_bestfit and LG15\_final have $\kappa_{\rm ion}$ higher than Fiducial at $M_{\star}<10^{6}\,\rm M_{\odot}$, due to their larger SFR, as shown in the bottom of Fig.~\ref{fig:pro_2ddis_halo_stellar_sfr}.
The bottom panel of Fig.~\ref{fig:pro_2ddis_stellar_sfr_ionflux} shows the 2-D distributions of $\kappa_{\rm ion}$ versus SFR at $z=7$ from all LG simulations. 
As in the top panel of the figure, LG20\_bestfit, LG20\_final, LG20\_dust\_bestfit and LG20\_dust\_final have $\kappa_{\rm ion}$ distributions similar to those of Fiducial.
Since LG15\_bestfit and LG15\_final do not have many galaxies with SFR $<10^{-6}\,\rm M_{\odot}\,yr^{-1}$ as shown at the bottom of Fig.~\ref{fig:pro_2ddis_halo_stellar_sfr}, their $\kappa_{\rm ion}$ show a smaller scatter than Fiducial at SFR $<10^{-6}\,\rm M_{\odot}\,yr^{-1}$, while their mean $\kappa_{\rm ion}$ versus SFR are consistent with the latter. 
Thus, we can conclude that the relation of $\kappa_{\rm ion}$ with $M_{\star}$ and SFR is insensitive to the choice of galaxy formation models. 
Further, the large scatter in the relations is due to the variations of star formation history and metallicity of different galaxies. 
}

\section{Conclusions and Discussions} 
\label{sec:conclu}
With the increasing number and quality of observations from the JWST telescope, it is now possible to put constrains on galaxy formation models during the Epoch of Reionization (EoR).
In this paper, we use the Markov Chain Monte Carlo (MCMC) method to fit galaxy formation models with UV luminosity functions (UVLF) observed by HST and JWST at $z \ge 6$, by combining the high-resolution {\it N}-body dark matter simulation {\sc Jiutian-300} and the semi-analytical galaxy formation model {\sc L-Galaxies} (LG). 
We then explore the galaxy properties and budget of ionizing photons during the EoR with the best-fit parameter values of the LG models.

We do three sets of MCMC calculations, the first one with {\sc L-Galaxies 2020} (LG20) by assuming no dust correction on the UV luminosity, the second one also with LG20 but with dust correction, and the last one with {\sc L-Galaxies 2015} (LG15) and no dust correction.
The MCMC fittings are done for 15 of the LG20 and 16 of the LG15 free parameters which govern the galaxy formation and evolution process. 
We then run seven sets of LG simulations to explore the galaxy properties and the budget of ionizing photons during the EoR.
Our Fiducial model adopts the parameter values from \cite{Henriques_2020}.
Three simulations have the best-fit parameter values from the three MCMC runs.
Since most free parameters are not well constrained by UVLF observations \cite[see also the discussions in ][]{Ma2023MNRAS}, we run three additional sets of LG simulations that adopt the best-fit values of parameters which are well limited by UVLF observations (i.e. $\alpha_{\rm H_{2}}$, $\beta_{\rm SF,\,burst}$, $\gamma_{\rm reinc}$ and $M_{\rm r.p.}$ for LG20, and $\alpha_{\rm SF}$ and $\eta_{\rm eject}$ for LG15), while the others are assumed to be the original ones in LG20 \citep{Henriques_2020} and LG15 \citep{Henriques_2015}. 

We find that with the fine-tuned sets of parameter values, both LG15 and LG20 can produce UVLF consistent with observations, while they still predict different budget of ionizing photons due to their different prescription for star formation.
Specifically, as dust correction reduces the bright end of the UVLF ($M_{\rm 1500,\, AB}<-20$), its inclusion in the MCMC calculation results in a higher star formation efficiency, which increases the stellar mass functions (SMF) at $M_{\star}>10^{9}\,\rm M_{\odot}$, and thus in a higher budget of ionizing photons.
The SMF from LG20 simulations that include dust correction in the MCMC calculation is more consistent than the other simulations with SMF observations at $z\ge6$ \citep{Stefanon2021, Navarro-Carrera2024}.
Because of the different prescription for the star formation process, the LG15 simulations have SFR and stellar mass higher than LG20 in low-mass halos ($M_{\rm halo}<10^{10}\,\rm M_{\odot}$), and as a consequence a larger SFR density, which predict $\ge 100\%$ more ionizing photons than LG20 models.
Additionally, the cold gas metallicity of galaxies with stellar mass $M_{\star}<10^{8}\,\rm M_{\odot}$ in LG15 is much higher than in LG20, which can explain why the dust correction reduces significantly the UVLF in \cite{Clay2015MNRAS}, who used LG15, but its effect is not very strong for the results obtained with LG20. 

In all seven LG simulations, the total number of ionizing photons after time integration along the star formation history of galaxies is linearly related to their stellar mass through a power-law relation with index $\sim 1$.
We also estimate the number of ionizing photons emitted per baryon ($\epsilon_{\rm ion}$) within stars, finding that $\epsilon_{\rm ion} \sim 10^{4}$ for all simulations, and that it slightly decreases with decreasing $z$.
Our estimated ionizing photon emission efficiency $\zeta_{\rm ion}$ is not sensitive to the stellar mass in galaxies nor to the absolute magnitude $M_{\rm 1500,\, AB}$ at $\lambda = 1500$\AA.
However, in models including the dust correction on the UV luminosity, the $\zeta_{\rm ion}$ is higher in the massive and bright galaxies than in the smaller ones. 
  
\begin{acknowledgments}
This work is supported by the National SKA Program of China (grant No. 2020SKA0110402), National Natural Science Foundation of China (Grant No. 12263002), and GZNU 2019 Special projects of training new academics and innovation exploration.
The {\sc Jiutian} simulations were conducted under the support of the science research grants from the China Manned Space Project with NO. CMS-CSST-2021-A03.
XW is supported by the National Natural Science Foundation of China (grant 12373009), the CAS Project for Young Scientists in Basic Research Grant No. YSBR-062, the Fundamental Research Funds for the Central Universities, the Xiaomi Young Talents Program, and the science research grant from the China Manned Space Project.
\end{acknowledgments}

\appendix

\section{Analysis of MCMC samples}
\label{app:mcmc}
{ Since the number of merger trees from {\it N}-body simulations is too huge to perform MCMC, \citealt{Henriques2013} proposed to randomly sample only a small fraction of the trees. 
To make sure the selected trees still include enough galaxies to properly compute the galaxy properties, such as the UVLF, the results of one full {\sc L-Galaxies} simulation are adopted to compute how many FOF halos  (i.e. $n_{i}$) are needed in each halo bin $i$ to match the full UVLF, by following (see Appendix~B of \citealt{Henriques2013}):
\begin{equation}
    \sum^{I}_{i=1} \frac{N_{i}^{2}}{n_{i}^{2}}\theta_{ij}=F^2\phi_{j}^2
\end{equation}
where $N_{i}$ is the number of FoF halos in the $i$th halo mass bin, $F$ is the fraction of the rms uncertainty of estimated UVLF over its true value, $\theta_{ij}$ is the average number of galaxies in a UV luminosity bin $j$ for the halos in a halo mass bin $i$, $\phi_{j} = \sum^{I}_{i=1} N_{i} \theta_{ij}$ is the UVLF in the UV luminosity bin $j$, and $I$ is the total number of halo mass bins. 
Note that $N_{i}$ is from the {\it N}-body dark matter simulation, $\theta_{ij}$ and $\phi_{j}$ are from the Fiducial {\sc L-Galaxies} simulation, and $F = 0.05$ to make sure the uncertainty of the estimated UVLF is smaller than the observational ones.

To make the MCMC efficient, we do not adopt the merger trees with the most massive but rare FOF halos, which might make the reproduced UVLF from random merger tree samples different from the one from the full {\sc L-Galaxies} simulation. 
We thus remove the UV luminosity bins if their differences are $>5\%$ when performing the MCMC. 
Since such differences only happen at the bright end of UVLF, we set a cut-off on $M_{\rm 1500,\, AB}$ as shown in Fig.~\ref{fig:uv_smf_multi}. 

We compute the MCMC chains with the Metropolis-Hastings method (see Appendix~B of \citealt{Henriques2013} for more details), with flat priors for the fitted parameters. 
The likelihood we adopted is the $\chi^{2}$ of UVLF, i.e. the $\chi^2 = \sum (\phi_{\rm obs} - \phi_{\rm sim})^2/\sigma_{\rm obs}^2$, where $\phi_{\rm obs}$ is the observed UVLF, $\phi_{\rm sim}$ is the simulated UVLF, and $\sigma_{\rm obs}$ is the error of observed UVLF. 
The proposed parameters in one step are accepted when $e^{-\chi^2/2} / e^{-\chi_{\rm pre}^2/2} >$ one random number (flat probability) between 0 and 1, where $\chi_{\rm pre}^2$ is the $\chi^2$ of previous step.
}
As mentioned in subsection~\ref{sec:mcmc_cal}, we do three sets of MCMC runs.
For each run, we have $10^5$ samples on the chains.
We use the python package GetDist \footnote{https://getdist.readthedocs.io/} to do the MCMC sample analysis \citep{Lewis2019arXiv}. 
As suggested by GetDist, during the analysis we drop the first 30\% of samples. 

\begin{figure*}
\centering
	\includegraphics[width=0.98\linewidth]{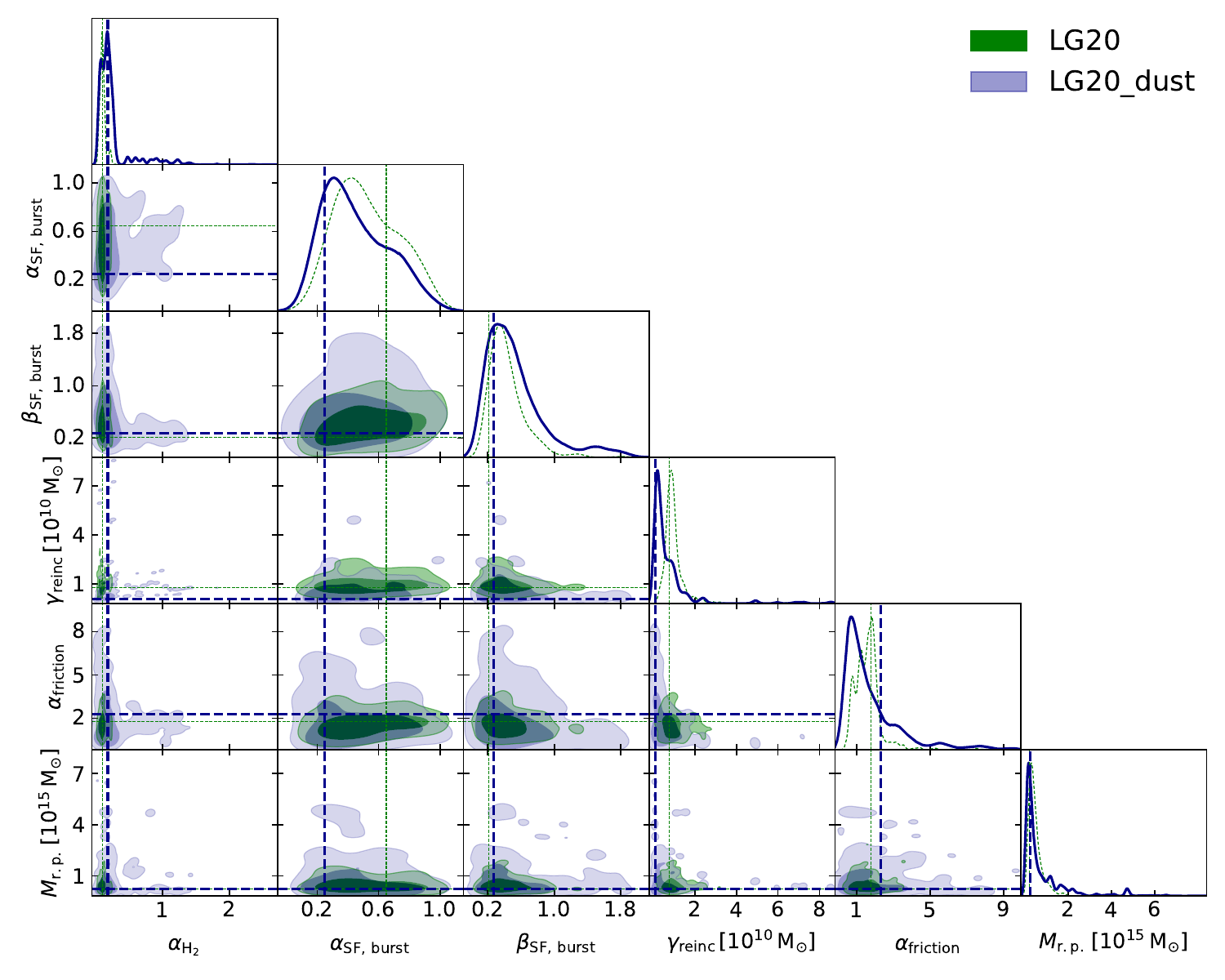}
    \caption{MCMC sample analysis for 6 free parameters in LG20 (green) without dust correction on UVLF, and LG20\_dust (dark blue) with dust correction on UVLF. 
    The first sub-plot in each column is the 1-D distribution of parameter samples, while the others are the 2-D distributions. The two contours for each color denote the 1-$\sigma$ ($68\%$) and 3-$\sigma$ ($95\%$) range of parameter samples. 
    The dashed lines denote the positions of the best-fit values of LG20 (green) and LG20\_dust (darkblue).
    }
    \label{fig:mcmc_LG20}
\end{figure*}
We find that most galaxy formation parameters are weakly limited by the UVLF observations (see the discussions in \citealt{Ma2023MNRAS}).
To make the results clearly visible, in Fig.~\ref{fig:mcmc_LG20} we show the 1-$\sigma$ ($68\%$) and 3-$\sigma$ ($95\%$) ranges of only 6 free parameters in LG20, including the case without dust correction on UVLF (named LG20) and the case with dust correction on UVLF (named LG20\_dust). 
4  of the 6 free parameters are well constrained by the UVLF observations at $z\ge 6$, i.e. $\alpha_{\rm H_{2}}$, $\beta_{\rm SF,\,burst}$, $\gamma_{\rm reinc}$ and $M_{\rm r.p.}$.
We note that both in LG20 and LG20\_dust, the original values of the 15 parameters in LG20 from \cite{Henriques_2020} are within the 1-$\sigma$ ranges limited by the UVLF at $z\ge 6$, i.e. the original parameter values of LG20 are already good at reproducing the UVLF observations at $z\ge 6$ (see the top panel of Fig.~\ref{fig:uv_smf_multi}).

\begin{figure*}
\centering
	\includegraphics[width=0.98\linewidth]{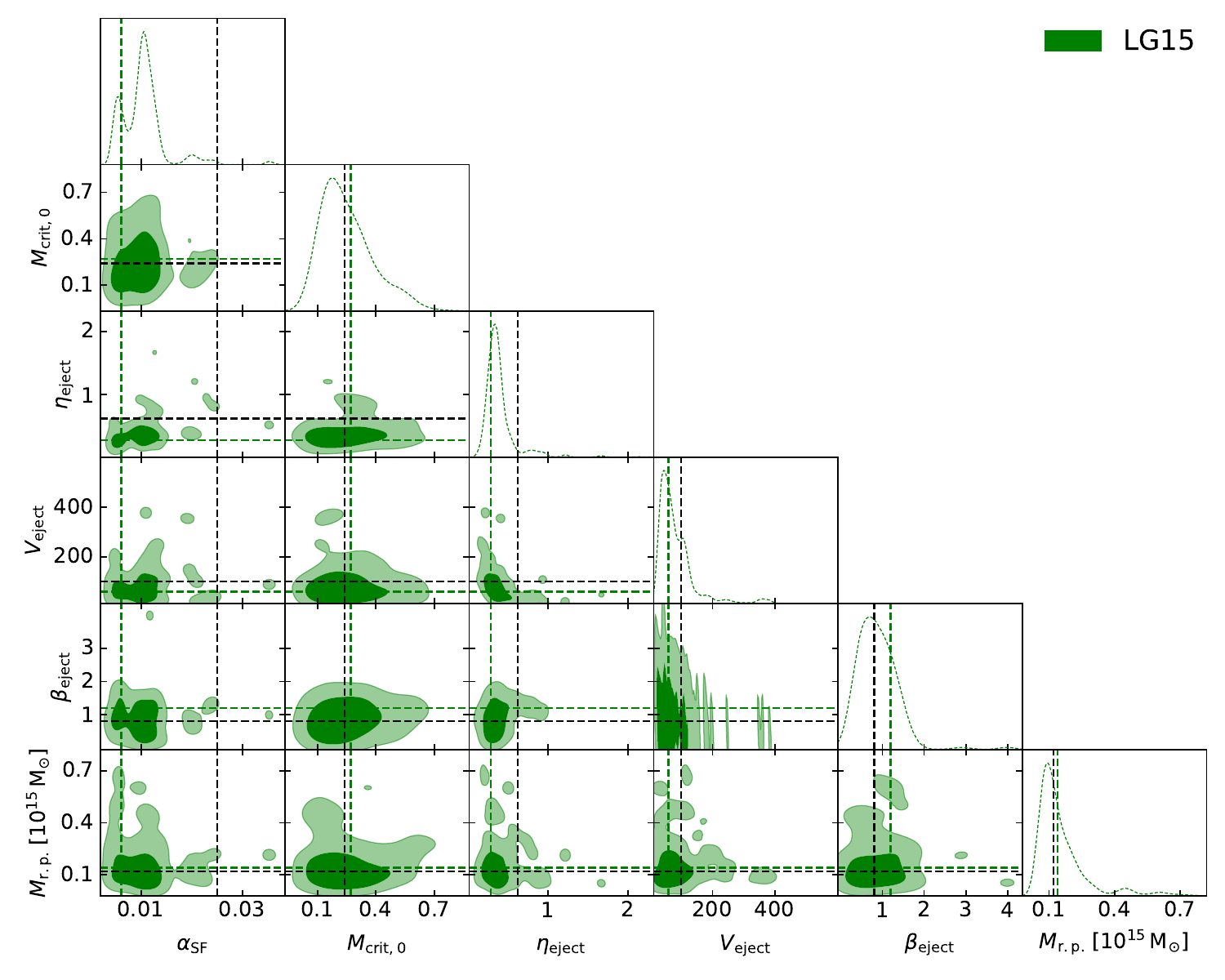}
    \caption{MCMC sample analysis for 6 free parameters in LG15 without dust correction on UVLF. 
    The first sub-plot in each volumn is the 1-D distribution of parameter samples, while the others are the 2-D distributions. The two contours denote the 1-$\sigma$ ($68\%$) and 3-$\sigma$ ($95\%$) range of parameter samples. 
    The dashed lines denote the positions of best-fit values in LG15 (green) and the original values in LG15 (black) from \cite{Henriques_2015}.
    }
    \label{fig:mcmc_LG15}
\end{figure*}
In Fig.~\ref{fig:mcmc_LG15}, we show the 1-$\sigma$ ($68\%$) and 3-$\sigma$ ($95\%$) ranges of 6 free parameters in LG15.
As a reference, we also show the positions of the original values in LG15 from \cite{Henriques_2015}.
We find that the original values of two parameters are beyond the 1-$\sigma$ range limited by the UVLF observations at $z\ge 6$, i.e. $\alpha_{\rm SF}$ and $\eta_{\rm eject}$. 
Note that our best-fit $\alpha_{\rm SF}$ is only $\sim 24\%$ of the original value. 

\bibliography{sample631}{}

\begin{thebibliography}{}
\expandafter\ifx\csname natexlab\endcsname\relax\def\natexlab#1{#1}\fi
\providecommand{\url}[1]{\href{#1}{#1}}
\providecommand{\dodoi}[1]{doi:~\href{http://doi.org/#1}{\nolinkurl{#1}}}
\providecommand{\doeprint}[1]{\href{http://ascl.net/#1}{\nolinkurl{http://ascl.net/#1}}}
\providecommand{\doarXiv}[1]{\href{https://arxiv.org/abs/#1}{\nolinkurl{https://arxiv.org/abs/#1}}}

\bibitem[{{Abdurashidova} {et~al.}(2022){Abdurashidova}, {Aguirre},
  {Alexander}, {Ali}, {Balfour}, {Beardsley}, {Bernardi}, {Billings}, {Bowman},
  {Bradley}, {Bull}, {Burba}, {Carey}, {Carilli}, {Cheng}, {DeBoer}, {Dexter},
  {de Lera Acedo}, {Dibblee-Barkman}, {Dillon}, {Ely}, {Ewall-Wice}, {Fagnoni},
  {Fritz}, {Furlanetto}, {Gale-Sides}, {Glendenning}, {Gorthi}, {Greig},
  {Grobbelaar}, {Halday}, {Hazelton}, {Hewitt}, {Hickish}, {Jacobs}, {Julius},
  {Kern}, {Kerrigan}, {Kittiwisit}, {Kohn}, {Kolopanis}, {Lanman}, {La Plante},
  {Lekalake}, {Lewis}, {Liu}, {MacMahon}, {Malan}, {Malgas}, {Maree},
  {Martinot}, {Matsetela}, {Mesinger}, {Molewa}, {Morales}, {Mosiane},
  {Murray}, {Neben}, {Nikolic}, {Nunhokee}, {Parsons}, {Patra}, {Pascua},
  {Pieterse}, {Pober}, {Razavi-Ghods}, {Ringuette}, {Robnett}, {Rosie}, {Sims},
  {Singh}, {Smith}, {Syce}, {Thyagarajan}, {Williams}, {Zheng}, \& {HERA
  Collaboration}}]{Abdurashidova2022}
{Abdurashidova}, Z., {Aguirre}, J.~E., {Alexander}, P., {et~al.} 2022, \apj,
  925, 221, \dodoi{10.3847/1538-4357/ac1c78}

\bibitem[{{Acharya} {et~al.}(2024{\natexlab{a}}){Acharya}, {Mertens}, {Ciardi},
  {Ghara}, {Koopmans}, \& {Zaroubi}}]{Acharya2024MNRAS}
{Acharya}, A., {Mertens}, F., {Ciardi}, B., {et~al.} 2024{\natexlab{a}},
  \mnras, 534, L30, \dodoi{10.1093/mnrasl/slae078}

\bibitem[{{Acharya} {et~al.}(2024{\natexlab{b}}){Acharya}, {Ma}, {Giri},
  {Ciardi}, {Ghara}, {Mellema}, {Zaroubi}, {Hothi}, {Iliev}, {Koopmans}, \&
  {Bianco}}]{Acharya2024arXiv}
{Acharya}, A., {Ma}, Q.-b., {Giri}, S.~K., {et~al.} 2024{\natexlab{b}}, arXiv
  e-prints, arXiv:2410.11620, \dodoi{10.48550/arXiv.2410.11620}

\bibitem[{{Adams} {et~al.}(2024){Adams}, {Conselice}, {Austin}, {Harvey},
  {Ferreira}, {Trussler}, {Juod{\v{z}}balis}, {Li}, {Windhorst}, {Cohen},
  {Jansen}, {Summers}, {Tompkins}, {Driver}, {Robotham}, {D'Silva}, {Yan},
  {Coe}, {Frye}, {Grogin}, {Koekemoer}, {Marshall}, {Pirzkal}, {Ryan},
  {Maksym}, {Rutkowski}, {Willmer}, {Hammel}, {Nonino}, {Bhatawdekar},
  {Wilkins}, {Bradley}, {Broadhurst}, {Cheng}, {Dole}, {Hathi}, \&
  {Zitrin}}]{Adams2024ApJ}
{Adams}, N.~J., {Conselice}, C.~J., {Austin}, D., {et~al.} 2024, \apj, 965,
  169, \dodoi{10.3847/1538-4357/ad2a7b}

\bibitem[{{Becker} {et~al.}(2015){Becker}, {Bolton}, {Madau}, {Pettini},
  {Ryan-Weber}, \& {Venemans}}]{Becker2015}
{Becker}, G.~D., {Bolton}, J.~S., {Madau}, P., {et~al.} 2015, \mnras, 447,
  3402, \dodoi{10.1093/mnras/stu2646}

\bibitem[{{Bhagwat} {et~al.}(2024){Bhagwat}, {Costa}, {Ciardi}, {Pakmor}, \&
  {Garaldi}}]{Bhagwat2024MNRAS}
{Bhagwat}, A., {Costa}, T., {Ciardi}, B., {Pakmor}, R., \& {Garaldi}, E. 2024,
  \mnras, 531, 3406, \dodoi{10.1093/mnras/stae1125}

\bibitem[{{Bosman} {et~al.}(2022){Bosman}, {Davies}, {Becker}, {Keating},
  {Davies}, {Zhu}, {Eilers}, {D'Odorico}, {Bian}, {Bischetti}, {Cristiani},
  {Fan}, {Farina}, {Haehnelt}, {Hennawi}, {Kulkarni}, {Mesinger}, {Meyer},
  {Onoue}, {Pallottini}, {Qin}, {Ryan-Weber}, {Schindler}, {Walter}, {Wang}, \&
  {Yang}}]{Bosman2022}
{Bosman}, S. E.~I., {Davies}, F.~B., {Becker}, G.~D., {et~al.} 2022, \mnras,
  514, 55, \dodoi{10.1093/mnras/stac1046}

\bibitem[{{Bouwens} {et~al.}(2020){Bouwens}, {Gonz{\'a}lez-L{\'o}pez},
  {Aravena}, {Decarli}, {Novak}, {Stefanon}, {Walter}, {Boogaard}, {Carilli},
  {Dudzevi{\v{c}}i{\={u}}t{\.{e}}}, {Smail}, {Daddi}, {da Cunha}, {Ivison},
  {Nanayakkara}, {Cortes}, {Cox}, {Inami}, {Oesch}, {Popping}, {Riechers}, {van
  der Werf}, {Weiss}, {Fudamoto}, \& {Wagg}}]{Bouwens2020}
{Bouwens}, R., {Gonz{\'a}lez-L{\'o}pez}, J., {Aravena}, M., {et~al.} 2020,
  \apj, 902, 112, \dodoi{10.3847/1538-4357/abb830}

\bibitem[{{Bouwens} {et~al.}(2015){Bouwens}, {Illingworth}, {Oesch}, {Trenti},
  {Labb{\'e}}, {Bradley}, {Carollo}, {van Dokkum}, {Gonzalez}, {Holwerda},
  {Franx}, {Spitler}, {Smit}, \& {Magee}}]{Bouwens2015ApJ}
{Bouwens}, R.~J., {Illingworth}, G.~D., {Oesch}, P.~A., {et~al.} 2015, \apj,
  803, 34, \dodoi{10.1088/0004-637X/803/1/34}

\bibitem[{{Bouwens} {et~al.}(2021){Bouwens}, {Oesch}, {Stefanon},
  {Illingworth}, {Labb{\'e}}, {Reddy}, {Atek}, {Montes}, {Naidu},
  {Nanayakkara}, {Nelson}, \& {Wilkins}}]{Bouwens2021}
{Bouwens}, R.~J., {Oesch}, P.~A., {Stefanon}, M., {et~al.} 2021, \aj, 162, 47,
  \dodoi{10.3847/1538-3881/abf83e}

\bibitem[{{Carniani} {et~al.}(2024){Carniani}, {Hainline}, {D'Eugenio},
  {Eisenstein}, {Jakobsen}, {Witstok}, {Johnson}, {Chevallard}, {Maiolino},
  {Helton}, {Willott}, {Robertson}, {Alberts}, {Arribas}, {Baker},
  {Bhatawdekar}, {Boyett}, {Bunker}, {Cameron}, {Cargile}, {Charlot}, {Curti},
  {Curtis-Lake}, {Egami}, {Giardino}, {Isaak}, {Ji}, {Jones}, {Kumari},
  {Maseda}, {Parlanti}, {P{\'e}rez-Gonz{\'a}lez}, {Rawle}, {Rieke}, {Rieke},
  {Del Pino}, {Saxena}, {Scholtz}, {Smit}, {Sun}, {Tacchella}, {{\"U}bler},
  {Venturi}, {Williams}, \& {Willmer}}]{Carniani2024Natur}
{Carniani}, S., {Hainline}, K., {D'Eugenio}, F., {et~al.} 2024, \nat, 633, 318,
  \dodoi{10.1038/s41586-024-07860-9}

\bibitem[{{Castellano} {et~al.}(2022){Castellano}, {Pentericci}, {Cupani},
  {Curtis-Lake}, {Vanzella}, {Amor{\'\i}n}, {Belfiori}, {Calabr{\`o}},
  {Carniani}, {Charlot}, {Chevallard}, {Dayal}, {Dickinson}, {Ferrara},
  {Fontana}, {Giallongo}, {Hutter}, {Merlin}, {Paris}, \&
  {Santini}}]{Castellano2022}
{Castellano}, M., {Pentericci}, L., {Cupani}, G., {et~al.} 2022, \aap, 662,
  A115, \dodoi{10.1051/0004-6361/202243348}

\bibitem[{{Clay} {et~al.}(2015){Clay}, {Thomas}, {Wilkins}, \&
  {Henriques}}]{Clay2015MNRAS}
{Clay}, S.~J., {Thomas}, P.~A., {Wilkins}, S.~M., \& {Henriques}, B. M.~B.
  2015, \mnras, 451, 2692, \dodoi{10.1093/mnras/stv818}

\bibitem[{{Croton} {et~al.}(2006){Croton}, {Springel}, {White}, {De Lucia},
  {Frenk}, {Gao}, {Jenkins}, {Kauffmann}, {Navarro}, \&
  {Yoshida}}]{Croton2006MNRAS}
{Croton}, D.~J., {Springel}, V., {White}, S. D.~M., {et~al.} 2006, \mnras, 365,
  11, \dodoi{10.1111/j.1365-2966.2005.09675.x}

\bibitem[{{Dayal} \& {Ferrara}(2018)}]{Dayal2018}
{Dayal}, P., \& {Ferrara}, A. 2018, \physrep, 780, 1,
  \dodoi{10.1016/j.physrep.2018.10.002}

\bibitem[{{Donnan} {et~al.}(2024){Donnan}, {McLure}, {Dunlop}, {McLeod},
  {Magee}, {Arellano-C{\'o}rdova}, {Barrufet}, {Begley}, {Bowler}, {Carnall},
  {Cullen}, {Ellis}, {Fontana}, {Illingworth}, {Grogin}, {Hamadouche},
  {Koekemoer}, {Liu}, {Mason}, {Santini}, \& {Stanton}}]{Donnan2024MNRAS}
{Donnan}, C.~T., {McLure}, R.~J., {Dunlop}, J.~S., {et~al.} 2024, \mnras, 533,
  3222, \dodoi{10.1093/mnras/stae2037}

\bibitem[{{Eldridge} {et~al.}(2017){Eldridge}, {Stanway}, {Xiao}, {McClelland},
  {Taylor}, {Ng}, {Greis}, \& {Bray}}]{Eldridge2017}
{Eldridge}, J.~J., {Stanway}, E.~R., {Xiao}, L., {et~al.} 2017, \pasa, 34,
  e058, \dodoi{10.1017/pasa.2017.51}

\bibitem[{{Esmerian} \& {Gnedin}(2021)}]{Esmerian2021}
{Esmerian}, C.~J., \& {Gnedin}, N.~Y. 2021, \apj, 910, 117,
  \dodoi{10.3847/1538-4357/abe869}

\bibitem[{{Fan} {et~al.}(2006){Fan}, {Carilli}, \& {Keating}}]{Fan2006}
{Fan}, X., {Carilli}, C.~L., \& {Keating}, B. 2006, \araa, 44, 415,
  \dodoi{10.1146/annurev.astro.44.051905.092514}

\bibitem[{{Finkelstein} {et~al.}(2024){Finkelstein}, {Leung}, {Bagley},
  {Dickinson}, {Ferguson}, {Papovich}, {Akins}, {Arrabal Haro}, {Dav{\'e}},
  {Dekel}, {Kartaltepe}, {Kocevski}, {Koekemoer}, {Pirzkal}, {Somerville},
  {Yung}, {Amor{\'\i}n}, {Backhaus}, {Behroozi}, {Bisigello}, {Bromm}, {Casey},
  {Ch{\'a}vez Ortiz}, {Cheng}, {Chworowsky}, {Cleri}, {Cooper}, {Davis}, {de la
  Vega}, {Elbaz}, {Franco}, {Fontana}, {Fujimoto}, {Giavalisco}, {Grogin},
  {Holwerda}, {Huertas-Company}, {Hirschmann}, {Iyer}, {Jogee}, {Jung},
  {Larson}, {Lucas}, {Mobasher}, {Morales}, {Morley}, {Mukherjee},
  {P{\'e}rez-Gonz{\'a}lez}, {Ravindranath}, {Rodighiero}, {Rowland},
  {Tacchella}, {Taylor}, {Trump}, \& {Wilkins}}]{Finkelstein2024ApJ}
{Finkelstein}, S.~L., {Leung}, G. C.~K., {Bagley}, M.~B., {et~al.} 2024, \apjl,
  969, L2, \dodoi{10.3847/2041-8213/ad4495}

\bibitem[{{Furlanetto} {et~al.}(2006){Furlanetto}, {Oh}, \&
  {Briggs}}]{Furlanetto2006}
{Furlanetto}, S.~R., {Oh}, S.~P., \& {Briggs}, F.~H. 2006, \physrep, 433, 181,
  \dodoi{10.1016/j.physrep.2006.08.002}

\bibitem[{{Gelli} {et~al.}(2024){Gelli}, {Mason}, \& {Hayward}}]{Gelli2024ApJ}
{Gelli}, V., {Mason}, C., \& {Hayward}, C.~C. 2024, \apj, 975, 192,
  \dodoi{10.3847/1538-4357/ad7b36}

\bibitem[{{Guo} {et~al.}(2011){Guo}, {White}, {Boylan-Kolchin}, {De Lucia},
  {Kauffmann}, {Lemson}, {Li}, {Springel}, \& {Weinmann}}]{Guo2011MNRAS}
{Guo}, Q., {White}, S., {Boylan-Kolchin}, M., {et~al.} 2011, \mnras, 413, 101,
  \dodoi{10.1111/j.1365-2966.2010.18114.x}

\bibitem[{{Harikane} {et~al.}(2024{\natexlab{a}}){Harikane}, {Nakajima},
  {Ouchi}, {Umeda}, {Isobe}, {Ono}, {Xu}, \& {Zhang}}]{Harikane2024ApJ}
{Harikane}, Y., {Nakajima}, K., {Ouchi}, M., {et~al.} 2024{\natexlab{a}}, \apj,
  960, 56, \dodoi{10.3847/1538-4357/ad0b7e}

\bibitem[{{Harikane} {et~al.}(2023){Harikane}, {Ouchi}, {Oguri}, {Ono},
  {Nakajima}, {Isobe}, {Umeda}, {Mawatari}, \& {Zhang}}]{Harikane2023ApJS}
{Harikane}, Y., {Ouchi}, M., {Oguri}, M., {et~al.} 2023, \apjs, 265, 5,
  \dodoi{10.3847/1538-4365/acaaa9}

\bibitem[{{Harikane} {et~al.}(2024{\natexlab{b}}){Harikane}, {Inoue}, {Ellis},
  {Ouchi}, {Nakazato}, {Yoshida}, {Ono}, {Sun}, {Sato}, {Fujimoto},
  {Kashikawa}, {McLeod}, {Perez-Gonzalez}, {Sawicki}, {Sugahara}, {Xu},
  {Yamanaka}, {Carnall}, {Cullen}, {Dunlop}, {Egami}, {Grogin}, {Isobe},
  {Koekemoer}, {Laporte}, {Lee}, {Magee}, {Matsuo}, {Matsuoka}, {Mawatari},
  {Nakajima}, {Nakane}, {Tamura}, {Umeda}, \& {Yanagisawa}}]{Harikane2024arXiv}
{Harikane}, Y., {Inoue}, A.~K., {Ellis}, R.~S., {et~al.} 2024{\natexlab{b}},
  arXiv e-prints, arXiv:2406.18352, \dodoi{10.48550/arXiv.2406.18352}

\bibitem[{{Henriques} {et~al.}(2015){Henriques}, {White}, {Thomas}, {Angulo},
  {Guo}, {Lemson}, {Springel}, \& {Overzier}}]{Henriques_2015}
{Henriques}, B. M.~B., {White}, S. D.~M., {Thomas}, P.~A., {et~al.} 2015,
  \mnras, 451, 2663, \dodoi{10.1093/mnras/stv705}

\bibitem[{{Henriques} {et~al.}(2013){Henriques}, {White}, {Thomas}, {Angulo},
  {Guo}, {Lemson}, \& {Springel}}]{Henriques2013}
---. 2013, \mnras, 431, 3373, \dodoi{10.1093/mnras/stt415}

\bibitem[{{Henriques} {et~al.}(2020){Henriques}, {Yates}, {Fu}, {Guo},
  {Kauffmann}, {Srisawat}, {Thomas}, \& {White}}]{Henriques_2020}
{Henriques}, B. M.~B., {Yates}, R.~M., {Fu}, J., {et~al.} 2020, \mnras, 491,
  5795, \dodoi{10.1093/mnras/stz3233}

\bibitem[{{Hutter} {et~al.}(2021){Hutter}, {Dayal}, {Yepes}, {Gottl{\"o}ber},
  {Legrand}, \& {Ucci}}]{Hutter2021}
{Hutter}, A., {Dayal}, P., {Yepes}, G., {et~al.} 2021, \mnras, 503, 3698,
  \dodoi{10.1093/mnras/stab602}

\bibitem[{{Kannan} {et~al.}(2022){Kannan}, {Garaldi}, {Smith}, {Pakmor},
  {Springel}, {Vogelsberger}, \& {Hernquist}}]{Kannan2022}
{Kannan}, R., {Garaldi}, E., {Smith}, A., {et~al.} 2022, \mnras, 511, 4005,
  \dodoi{10.1093/mnras/stab3710}

\bibitem[{{Koopmans} {et~al.}(2015){Koopmans}, {Pritchard}, {Mellema},
  {Aguirre}, {Ahn}, {Barkana}, {van Bemmel}, {Bernardi}, {Bonaldi}, {Briggs},
  {de Bruyn}, {Chang}, {Chapman}, {Chen}, {Ciardi}, {Dayal}, {Ferrara},
  {Fialkov}, {Fiore}, {Ichiki}, {Illiev}, {Inoue}, {Jelic}, {Jones}, {Lazio},
  {Maio}, {Majumdar}, {Mack}, {Mesinger}, {Morales}, {Parsons}, {Pen},
  {Santos}, {Schneider}, {Semelin}, {de Souza}, {Subrahmanyan}, {Takeuchi},
  {Vedantham}, {Wagg}, {Webster}, {Wyithe}, {Datta}, \& {Trott}}]{Koopmans2015}
{Koopmans}, L., {Pritchard}, J., {Mellema}, G., {et~al.} 2015, in Advancing
  Astrophysics with the Square Kilometre Array (AASKA14), 1,
  \dodoi{10.22323/1.215.0001}

\bibitem[{{Lewis}(2019)}]{Lewis2019arXiv}
{Lewis}, A. 2019, arXiv e-prints, arXiv:1910.13970,
  \dodoi{10.48550/arXiv.1910.13970}

\bibitem[{{Liu} {et~al.}(2024){Liu}, {Ma}, {Han}, \& {Luo}}]{Liu2024ApJ}
{Liu}, P., {Ma}, Q., {Han}, Y., \& {Luo}, R. 2024, \apj, 968, 13,
  \dodoi{10.3847/1538-4357/ad41e1}

\bibitem[{{Ma} {et~al.}(2017){Ma}, {Maio}, {Ciardi}, \&
  {Salvaterra}}]{Ma2017MNRAS}
{Ma}, Q., {Maio}, U., {Ciardi}, B., \& {Salvaterra}, R. 2017, \mnras, 472,
  3532, \dodoi{10.1093/mnras/stx1839}

\bibitem[{{Ma} {et~al.}(2023){Ma}, {Ghara}, {Ciardi}, {Iliev}, {Koopmans},
  {Mellema}, {Mondal}, \& {Zaroubi}}]{Ma2023MNRAS}
{Ma}, Q.-B., {Ghara}, R., {Ciardi}, B., {et~al.} 2023, \mnras, 522, 3284,
  \dodoi{10.1093/mnras/stad1203}

\bibitem[{{Ma} {et~al.}(2018){Ma}, {Hopkins}, {Garrison-Kimmel},
  {Faucher-Gigu{\`e}re}, {Quataert}, {Boylan-Kolchin}, {Hayward}, {Feldmann},
  \& {Kere{\v{s}}}}]{MaX2018}
{Ma}, X., {Hopkins}, P.~F., {Garrison-Kimmel}, S., {et~al.} 2018, \mnras, 478,
  1694, \dodoi{10.1093/mnras/sty1024}

\bibitem[{{Mertens} {et~al.}(2020){Mertens}, {Mevius}, {Koopmans}, {Offringa},
  {Mellema}, {Zaroubi}, {Brentjens}, {Gan}, {Gehlot}, {Pandey}, {Sardarabadi},
  {Vedantham}, {Yatawatta}, {Asad}, {Ciardi}, {Chapman}, {Gazagnes}, {Ghara},
  {Ghosh}, {Giri}, {Iliev}, {Jeli{\'c}}, {Kooistra}, {Mondal}, {Schaye}, \&
  {Silva}}]{Mertens2020}
{Mertens}, F.~G., {Mevius}, M., {Koopmans}, L.~V.~E., {et~al.} 2020, \mnras,
  493, 1662, \dodoi{10.1093/mnras/staa327}

\bibitem[{{Mutch} {et~al.}(2016){Mutch}, {Geil}, {Poole}, {Angel}, {Duffy},
  {Mesinger}, \& {Wyithe}}]{Mutch2016}
{Mutch}, S.~J., {Geil}, P.~M., {Poole}, G.~B., {et~al.} 2016, \mnras, 462, 250,
  \dodoi{10.1093/mnras/stw1506}

\bibitem[{{Nakajima} {et~al.}(2023){Nakajima}, {Ouchi}, {Isobe}, {Harikane},
  {Zhang}, {Ono}, {Umeda}, \& {Oguri}}]{Nakajima2023ApJS}
{Nakajima}, K., {Ouchi}, M., {Isobe}, Y., {et~al.} 2023, \apjs, 269, 33,
  \dodoi{10.3847/1538-4365/acd556}

\bibitem[{{Navarro-Carrera} {et~al.}(2024){Navarro-Carrera}, {Rinaldi},
  {Caputi}, {Iani}, {Kokorev}, \& {van Mierlo}}]{Navarro-Carrera2024}
{Navarro-Carrera}, R., {Rinaldi}, P., {Caputi}, K.~I., {et~al.} 2024, \apj,
  961, 207, \dodoi{10.3847/1538-4357/ad0df6}

\bibitem[{{Nikoli{\'c}} {et~al.}(2024){Nikoli{\'c}}, {Mesinger}, {Davies}, \&
  {Prelogovi{\'c}}}]{Nikolic2024arXiv}
{Nikoli{\'c}}, I., {Mesinger}, A., {Davies}, J.~E., \& {Prelogovi{\'c}}, D.
  2024, arXiv e-prints, arXiv:2406.15237, \dodoi{10.48550/arXiv.2406.15237}

\bibitem[{{Ocvirk} {et~al.}(2020){Ocvirk}, {Aubert}, {Sorce}, {Shapiro},
  {Deparis}, {Dawoodbhoy}, {Lewis}, {Teyssier}, {Yepes}, {Gottl{\"o}ber},
  {Ahn}, {Iliev}, \& {Hoffman}}]{Ocvirk2020MNRAS}
{Ocvirk}, P., {Aubert}, D., {Sorce}, J.~G., {et~al.} 2020, \mnras, 496, 4087,
  \dodoi{10.1093/mnras/staa1266}

\bibitem[{{Park} {et~al.}(2019){Park}, {Mesinger}, {Greig}, \&
  {Gillet}}]{Park2019MNRAS}
{Park}, J., {Mesinger}, A., {Greig}, B., \& {Gillet}, N. 2019, \mnras, 484,
  933, \dodoi{10.1093/mnras/stz032}

\bibitem[{{Planck Collaboration} {et~al.}(2020){Planck Collaboration},
  {Aghanim}, {Akrami}, {Ashdown}, {Aumont}, {Baccigalupi}, {Ballardini},
  {Banday}, {Barreiro}, {Bartolo}, {Basak}, {Battye}, {Benabed}, {Bernard},
  {Bersanelli}, {Bielewicz}, {Bock}, {Bond}, {Borrill}, {Bouchet}, {Boulanger},
  {Bucher}, {Burigana}, {Butler}, {Calabrese}, {Cardoso}, {Carron},
  {Challinor}, {Chiang}, {Chluba}, {Colombo}, {Combet}, {Contreras}, {Crill},
  {Cuttaia}, {de Bernardis}, {de Zotti}, {Delabrouille}, {Delouis}, {Di
  Valentino}, {Diego}, {Dor{\'e}}, {Douspis}, {Ducout}, {Dupac}, {Dusini},
  {Efstathiou}, {Elsner}, {En{\ss}lin}, {Eriksen}, {Fantaye}, {Farhang},
  {Fergusson}, {Fernandez-Cobos}, {Finelli}, {Forastieri}, {Frailis},
  {Fraisse}, {Franceschi}, {Frolov}, {Galeotta}, {Galli}, {Ganga},
  {G{\'e}nova-Santos}, {Gerbino}, {Ghosh}, {Gonz{\'a}lez-Nuevo}, {G{\'o}rski},
  {Gratton}, {Gruppuso}, {Gudmundsson}, {Hamann}, {Handley}, {Hansen},
  {Herranz}, {Hildebrandt}, {Hivon}, {Huang}, {Jaffe}, {Jones}, {Karakci},
  {Keih{\"a}nen}, {Keskitalo}, {Kiiveri}, {Kim}, {Kisner}, {Knox},
  {Krachmalnicoff}, {Kunz}, {Kurki-Suonio}, {Lagache}, {Lamarre}, {Lasenby},
  {Lattanzi}, {Lawrence}, {Le Jeune}, {Lemos}, {Lesgourgues}, {Levrier},
  {Lewis}, {Liguori}, {Lilje}, {Lilley}, {Lindholm}, {L{\'o}pez-Caniego},
  {Lubin}, {Ma}, {Mac{\'\i}as-P{\'e}rez}, {Maggio}, {Maino}, {Mandolesi},
  {Mangilli}, {Marcos-Caballero}, {Maris}, {Martin}, {Martinelli},
  {Mart{\'\i}nez-Gonz{\'a}lez}, {Matarrese}, {Mauri}, {McEwen}, {Meinhold},
  {Melchiorri}, {Mennella}, {Migliaccio}, {Millea}, {Mitra},
  {Miville-Desch{\^e}nes}, {Molinari}, {Montier}, {Morgante}, {Moss}, {Natoli},
  {N{\o}rgaard-Nielsen}, {Pagano}, {Paoletti}, {Partridge}, {Patanchon},
  {Peiris}, {Perrotta}, {Pettorino}, {Piacentini}, {Polastri}, {Polenta},
  {Puget}, {Rachen}, {Reinecke}, {Remazeilles}, {Renzi}, {Rocha}, {Rosset},
  {Roudier}, {Rubi{\~n}o-Mart{\'\i}n}, {Ruiz-Granados}, {Salvati}, {Sandri},
  {Savelainen}, {Scott}, {Shellard}, {Sirignano}, {Sirri}, {Spencer},
  {Sunyaev}, {Suur-Uski}, {Tauber}, {Tavagnacco}, {Tenti}, {Toffolatti},
  {Tomasi}, {Trombetti}, {Valenziano}, {Valiviita}, {Van Tent}, {Vibert},
  {Vielva}, {Villa}, {Vittorio}, {Wandelt}, {Wehus}, {White}, {White},
  {Zacchei}, \& {Zonca}}]{Planck2020A&A}
{Planck Collaboration}, {Aghanim}, N., {Akrami}, Y., {et~al.} 2020, \aap, 641,
  A6, \dodoi{10.1051/0004-6361/201833910}

\bibitem[{{Rosdahl} {et~al.}(2018){Rosdahl}, {Katz}, {Blaizot}, {Kimm},
  {Michel-Dansac}, {Garel}, {Haehnelt}, {Ocvirk}, \& {Teyssier}}]{Rosdahl2018}
{Rosdahl}, J., {Katz}, H., {Blaizot}, J., {et~al.} 2018, \mnras, 479, 994,
  \dodoi{10.1093/mnras/sty1655}

\bibitem[{{Simmonds} {et~al.}(2024){Simmonds}, {Tacchella}, {Hainline},
  {Johnson}, {Pusk{\'a}s}, {Robertson}, {Baker}, {Bhatawdekar}, {Boyett},
  {Bunker}, {Cargile}, {Carniani}, {Chevallard}, {Curti}, {Curtis-Lake}, {Ji},
  {Jones}, {Kumari}, {Laseter}, {Maiolino}, {Maseda}, {Rinaldi}, {Stoffers},
  {{\"U}bler}, {Villanueva}, {Williams}, {Willott}, {Witstok}, \&
  {Zhu}}]{Simmonds2024MNRAS}
{Simmonds}, C., {Tacchella}, S., {Hainline}, K., {et~al.} 2024, \mnras, 535,
  2998, \dodoi{10.1093/mnras/stae2537}

\bibitem[{{Springel} {et~al.}(2021){Springel}, {Pakmor}, {Zier}, \&
  {Reinecke}}]{Springel2021}
{Springel}, V., {Pakmor}, R., {Zier}, O., \& {Reinecke}, M. 2021, \mnras, 506,
  2871, \dodoi{10.1093/mnras/stab1855}

\bibitem[{{Springel} {et~al.}(2001){Springel}, {White}, {Tormen}, \&
  {Kauffmann}}]{Springel2001}
{Springel}, V., {White}, S. D.~M., {Tormen}, G., \& {Kauffmann}, G. 2001,
  \mnras, 328, 726, \dodoi{10.1046/j.1365-8711.2001.04912.x}

\bibitem[{{Springel} {et~al.}(2005){Springel}, {White}, {Jenkins}, {Frenk},
  {Yoshida}, {Gao}, {Navarro}, {Thacker}, {Croton}, {Helly}, {Peacock}, {Cole},
  {Thomas}, {Couchman}, {Evrard}, {Colberg}, \& {Pearce}}]{Springel2005}
{Springel}, V., {White}, S. D.~M., {Jenkins}, A., {et~al.} 2005, \nat, 435,
  629, \dodoi{10.1038/nature03597}

\bibitem[{{Stanway} \& {Eldridge}(2018)}]{Stanway2018MNRAS}
{Stanway}, E.~R., \& {Eldridge}, J.~J. 2018, \mnras, 479, 75,
  \dodoi{10.1093/mnras/sty1353}

\bibitem[{{Stefanon} {et~al.}(2021){Stefanon}, {Bouwens}, {Labb{\'e}},
  {Illingworth}, {Gonzalez}, \& {Oesch}}]{Stefanon2021}
{Stefanon}, M., {Bouwens}, R.~J., {Labb{\'e}}, I., {et~al.} 2021, \apj, 922,
  29, \dodoi{10.3847/1538-4357/ac1bb6}

\bibitem[{{Tang} {et~al.}(2023){Tang}, {Stark}, {Chen}, {Mason}, {Topping},
  {Endsley}, {Senchyna}, {Plat}, {Lu}, {Whitler}, {Robertson}, \&
  {Charlot}}]{Tang2023}
{Tang}, M., {Stark}, D.~P., {Chen}, Z., {et~al.} 2023, \mnras, 526, 1657,
  \dodoi{10.1093/mnras/stad2763}

\bibitem[{{Tinker} {et~al.}(2008){Tinker}, {Kravtsov}, {Klypin}, {Abazajian},
  {Warren}, {Yepes}, {Gottl{\"o}ber}, \& {Holz}}]{Tinker2008}
{Tinker}, J., {Kravtsov}, A.~V., {Klypin}, A., {et~al.} 2008, \apj, 688, 709,
  \dodoi{10.1086/591439}

\bibitem[{{Trott} {et~al.}(2020){Trott}, {Jordan}, {Midgley}, {Barry}, {Greig},
  {Pindor}, {Cook}, {Sleap}, {Tingay}, {Ung}, {Hancock}, {Williams}, {Bowman},
  {Byrne}, {Chokshi}, {Hazelton}, {Hasegawa}, {Jacobs}, {Joseph}, {Li}, {Line},
  {Lynch}, {McKinley}, {Mitchell}, {Morales}, {Ouchi}, {Pober}, {Rahimi},
  {Takahashi}, {Wayth}, {Webster}, {Wilensky}, {Wyithe}, {Yoshiura}, {Zhang},
  \& {Zheng}}]{Trott2020}
{Trott}, C.~M., {Jordan}, C.~H., {Midgley}, S., {et~al.} 2020, \mnras, 493,
  4711, \dodoi{10.1093/mnras/staa414}

\bibitem[{{Vani} {et~al.}(2024){Vani}, {Ayromlou}, {Kauffmann}, \&
  {Springel}}]{Vani2024}
{Vani}, A., {Ayromlou}, M., {Kauffmann}, G., \& {Springel}, V. 2024, arXiv
  e-prints, arXiv:2408.00824, \dodoi{10.48550/arXiv.2408.00824}

\bibitem[{{Wang} {et~al.}(2024){Wang}, {Cheng}, {Ge}, {Meng}, {Daddi}, {Yan},
  {Ji}, {Jin}, {Jones}, {Malkan}, {Arrabal Haro}, {Brammer}, {Oguri}, {Hou}, \&
  {Zhang}}]{Wang2024ApJ}
{Wang}, X., {Cheng}, C., {Ge}, J., {et~al.} 2024, \apjl, 967, L42,
  \dodoi{10.3847/2041-8213/ad4ced}

\bibitem[{{Wilkins} {et~al.}(2016){Wilkins}, {Feng}, {Di-Matteo}, {Croft},
  {Stanway}, {Bouwens}, \& {Thomas}}]{Wilkins2016MNRAS}
{Wilkins}, S.~M., {Feng}, Y., {Di-Matteo}, T., {et~al.} 2016, \mnras, 458, L6,
  \dodoi{10.1093/mnrasl/slw007}

\bibitem[{{Zhang} {et~al.}(2022){Zhang}, {Shan}, {Gu}, {Zheng}, {Xu}, {Yue},
  {Liu}, {Zhu}, \& {Guo}}]{Zhang2022MNRAS}
{Zhang}, Z., {Shan}, H., {Gu}, J., {et~al.} 2022, \mnras, 516, 1573,
  \dodoi{10.1093/mnras/stac2208}

\end{thebibliography}
\bibliographystyle{aasjournal}

\end{document}